\documentclass[usegraphicx,useAMS,usenatbib]{mn2e}
\usepackage{times}
\usepackage{rotating}
\usepackage{subfigure}

\title[Coronal super-saturation in M-dwarfs]
 {Investigating coronal saturation and super-saturation in fast-rotating M-dwarf stars}

\author[R.D. Jeffries et al.]
  {R.D.~Jeffries$^1$, R.J.~Jackson$^1$, K.R.~Briggs$^2$, P.A.~Evans$^3$ and J.P.~Pye$^3$\\
  $^1$ Astrophysics Group, Research Institute for the Environment, Physical
  Sciences and Applied Mathematics, Keele University, Staffordshire ST5
  5BG, UK\\
  $^2$ ETH Zurich, Institute of Astronomy, 8093 Zurich, Switzerland \\
  $^3$ Department of Physics and Astronomy, University of Leicester,
  University Road, Leicester, LE1 7RH, UK}
\date{Accepted October 11th 2010}

\pagerange{\pageref{firstpage}--\pageref{lastpage}} \pubyear{2010}

\def\LaTeX{L\kern-.36em\raise.3ex\hbox{a}\kern-.15em
    T\kern-.1667em\lower.7ex\hbox{E}\kern-.125emX}

\begin{document}
\label{firstpage}
\maketitle

\begin{abstract}
At fast rotation rates the coronal activity of G- and K-type stars has
been observed to ``saturate'' and then decline again at even faster
rotation rates -- a phenomenon dubbed ``super-saturation''. In this
paper we investigate coronal activity in fast-rotating M-dwarfs using
deep {\it XMM-Newton} observations of 97 low-mass
stars of known rotation period in the young
open cluster NGC~2547, and combine these with published X-ray surveys
of low-mass field and cluster stars of known rotation period. Like G-
and K-dwarfs, we find that M-dwarfs exhibit increasing coronal activity
with decreasing Rossby number $N_R$, the ratio of period to convective
turnover time, and that activity saturates at $L_x/L_{\rm bol} \simeq
10^{-3}$ for $\log N_R < -0.8$. However, super-saturation is not
convincingly displayed by M-dwarfs, despite the presence of many
objects in our sample with $\log N_{R} < -1.8$, where 
super-saturation is observed to occur in higher mass
stars. Instead, it appears that a short rotation period is the primary
predictor of super-saturation; $P\leq 0.3$\,d for K-dwarfs and perhaps
$P\leq 0.2$\,d for M-dwarfs. These observations favour the
``centrifugal stripping'' model for super-saturation, where coronal
structures are forced open or become radiatively unstable as the
Keplerian co-rotation radius moves inside the X-ray emitting coronal
volume.
\end{abstract}


\section{Introduction}

X-ray emission from the hot coronae of photospherically cool stars
arises from magnetically confined and heated plasma with temperatures
in excess of $10^{6}$\,K (see the review by G\"udel 2004).  X-ray
observations of cool stars with convective envelopes
have long since shown that X-ray activity increases with rotation rate
and has led to the paradigm that a magnetic dynamo process
produces and maintains the required magnetic
fields, in analogy to processes observed to occur in the Sun
(e.g. Pallavicini et al. 1981; Mangeney \& Praderie 1984).  The
influence of a regenerative dynamo is supported by
observations of coronal activity in stars at a range of masses.  These
demonstrate, in accordance with expectations from dynamo models, that
rotation is not the only important parameter. The best correlations are
found between increasing magnetic activity and the inverse of the
Rossby number (e.g. Noyes et al. 1984; Dobson \& Radick 1989; Pizzolato
et al. 2003), where Rossby number is $P/\tau_c$, the ratio of stellar
rotation period, $P$, to the convective turnover time, $\tau_c$. According
to stellar models, $\tau_c$ increases towards lower masses
(e.g. Gilliland 1986; Kim \& Demarque 1996) and hence, for a given
rotation period, a K-dwarf has a smaller Rossby number and is more
magnetically active than a G-dwarf, where magnetic activity is
expressed in a form normalised by the bolometric luminosity (e.g. the
ratio of coronal X-ray to bolometric luminosity, $L_x/L_{\rm bol}$).

Many young, cool stars have not lost significant angular momentum and
rotate much more rapidly than the Sun, leading to higher coronal X-ray
luminosities by several orders of magnitude. At fast rotation rates, magnetic
activity (as measured in the corona and chromosphere) appears to
``saturate'' (e.g. Vilhu \& Walter 1987; Stauffer et
al. 1994). Saturated coronal activity is manifested as a plateau of
$L_x/L_{\rm bol} \simeq 10^{-3}$ in G-dwarfs with rotation periods less
than about 3 days. In lower mass stars the plateau in coronal activity
is at a similar level but the rotation period at which it first occurs
is longer. A unified picture emerges by plotting X-ray activity against
Rossby number (St{\c e}pie\'n 1994; Randich et al. 1996; Patten \& Simon 1996;
Pizzolato et al. 2003), such that coronal
saturation occurs at Rossby numbers of about 0.1 in stars of spectral
types G, K and M.  This unification in terms of a parameter associated
with dynamo efficiency suggests that coronal saturation reflects a
saturation of the dynamo itself (Vilhu \& Walter 1987), although other
explanations, such as a redistribution of radiative losses to other
wavelengths (Doyle 1996) or changes in magnetic topology with fast
rotation (Solanki, Motamen \& Keppens 1997; Jardine \& Unruh 1999;
Ryan, Neukirch \& Jardine 2005), must also be considered.

At rotation rates $\sim 5$ times faster than required for saturation it
appears that coronal activity turns down again -- a phenomenon dubbed
``super-saturation'' (Prosser et al. 1996). Examples of super-saturated
G- and K-stars, with $L_x/L_{\rm bol} \simeq 10^{-3.5}$, have been
found in a number of young open clusters (Stauffer et al. 1994, 1997;
Patten \& Simon 1996; Randich 1998; Jeffries et al. 2006),
among the fast rotating components of contact W~UMa binaries (Cruddace
\& Dupree 1984; St{\c e}pie\'n, Schmitt \& Voges 2001) and has been
suggested as the reason for lower X-ray activity in the
fastest rotating, very young pre main-sequence stars (Stassun et
al. 2004; Preibisch et al. 2005; Dahm et al. 2007). Possible
explanations for
coronal super-saturation include negative feedback in the dynamo
(Kitchatinov, R\"udiger \& K\"uker 1994), decreasing coverage by
active regions (St{\c e}pie\'n et al. 2001), reorganisation of the coronal
magnetic field (Solanki et al. 1997) or centrifugal stripping of the
corona (Jardine 2004).

An important test of these ideas is to look at the coronal properties
of fast rotators across a wide range of masses. In particular, it
is vital to gauge whether saturation and super-saturation occur at
fixed values of rotation period or Rossby number. This would
illuminate which physical mechanisms are responsible. One of the
principal gaps in our knowledge is the behaviour of coronal emission
for ultra-fast rotating M-dwarfs. These have larger convection zones
(as a fraction of the star), longer convective turnover times and
hence smaller Rossby numbers at a given rotation period than G- and K-dwarfs.
Of course M-dwarfs also have much lower
bolometric luminosities than G- or K-dwarfs and so, at a given magnetic
activity level, they are harder to observe in the young open clusters
where the majority of ultra-fast rotators are found.

The most comprehensive work so far was by James et al. (2000) using
X-ray observations for a small, inhomogeneous sample of fast rotating low-mass stars
from the field and open clusters. 
They found that, like G- and K-type stars, M-dwarfs
with periods below $\sim 6$ days and Rossby numbers below 0.1 showed
saturated levels of X-ray emission with $L_x/L_{\rm bol} \simeq
10^{-3}$. They also claimed tentative evidence for super-saturation in
the fastest rotating M-dwarfs, with periods of 0.2--0.3 days. 

In this paper we re-visit the question of saturation and
super-saturation of coronal emission in fast-rotating M-dwarfs. We analyse
new, deep {\it XMM-Newton} observations of a large,
homogeneous sample of rapidly rotating M-dwarfs identified in the open
cluster NGC~2547 by Irwin et al. (2008). NGC~2547 has an age of
35--38\,Myr (Jeffries \& Oliveira 2005; Naylor \& Jeffries 2006) and a
rich population of low-mass stars (Jeffries et al. 2004). However, the
cluster is at at $\sim 400$\,pc, and while previous X-ray observations
with {\it ROSAT} (Jeffries \& Tolley 1998) and {\it XMM-Newton}
(Jeffries et al. 2006) demonstrated an X-ray active low-mass
population, they were insufficiently sensitive to probe the coronal
activity of its M-dwarfs in any detail.

In section~2 we describe the new {\it XMM-Newton} observations of
NGC~2547 and the identification of X-ray sources with rapidly rotating
M-dwarfs from the Irwin et al. (2008) catalogue.  In section~3 we
combine X-ray and optical data, estimate coronal activity levels and
examine the evidence for coronal saturation and super-saturation using
a homogeneous sample of fast rotating M-dwarfs several times larger than considered
by James et al. (2000). In section 4 we compare our results to those
compiled in the literature for G- and K-stars and for other samples of
M-dwarfs. In section 5 we discuss our results in the context of
competing models for saturation/super-saturation and our conclusions
are summarised in section 6.

\section{XMM-Newton Observations of NGC~2547}
\label{observations}

NGC 2547 was observed with {\it XMM-Newton} between UT 22:30:03 on 12
November 2007 and UT 09:30:03 on 14 November 2007 using the European
Photon Imaging Counter (EPIC) instrument,
for a nominal exposure time of 125.8\,ks (Observation ID 0501790101).  
The two EPIC-MOS cameras and
the EPIC-PN camera were operated in full frame mode (Turner et
al. 2001; Str\"uder et al. 2001),
using the medium filter to reject optical light. The nominal
pointing position of the observation was RA\,$=08$h\,10m\,06.1s,
Dec\,$=-49$d\,15m\,42.9s (J2000.0).

\subsection{Source Detection}

Version 7.1 of the {\it XMM-Newton} Science Analysis System was used
for the initial data reduction and source detection.
Data from the three
cameras were individually screened for high background periods and
these time intervals were excluded from all subsequent analysis. Observation
intervals were excluded where the total count rate 
for events with energies $>10$\,keV,
exceeded 0.35\,s$^{-1}$ and 1.0\,s$^{-1}$ for the MOS and PN detectors
respectively.  The remaining useful exposure times were 107.3\,ks,
104.8\,ks and 87.4\,ks for the MOS1, MOS2 and PN cameras respectively,
which can be compared with the equivalent exposure times of 29.0\,ks, 29.4\,ks
and 13.6\,ks in the less sensitive observation analysed by Jeffries et
al. (2006).

Images were created using the {\sc evselect} task and a spatial
sampling of 2 arcseconds per pixel. The event lists were filtered to
exclude anomalous pixel patterns and edge effects by including only
those events with ``pattern''\,$\leq 12$ for the MOS detectors and
$\leq 4$ for the PN detectors.  The contrast between background and
source events was also increased by retaining only those events in
channels corresponding to energies of 0.3--3\,keV.

The {\sc edetect\_chain} task was used to find sources with a combined
maximum log likelihood value greater than 10 
(approximately equivalent to $-\ln p$, where
$p$ is the probability that the ``source'' is due to a background
fluctuation), in all three instruments combined, over
the 0.3--3\,keV energy range.  We expect 1-2 spurious X-ray detections
at this level of significance, though they would be highly unlikely to
correlate with NGC~2547 members, so will not hamper any analysis in
this paper.  The individual images from each instrument were
source-searched first to confirm there were no systematic
differences in the astrometry of the brightest sources.  Count rates in
each detector were determined using vignetting-corrected
exposure maps created within the same task. In addition, count rates
were determined for each source in the 0.3--1.0\,keV and 1.0--3.0\,keV
bands separately, in order to form a hardness ratio.  
A total of 323 significant X-ray sources were found.
Some of these only have count rates measured in a subset of the three
instruments because they fell in gaps between detectors, on hot pixels
or lay outside the field of view. In addition we decided only to retain
count rates for analysis if they had a signal-to-noise ratio greater
than 3, which resulted in the removal of three sources from our list.

To check the EPIC astrometric solution, we cross-correlated all
the brightest X-rays sources (those detected with a maximum log likelihood
greater than 100) against a list of photometrically selected NGC~2547 members
compiled by Naylor et al. (2002 -- their Table~6), which is 
based on D'Antona \& Mazzitelli (1997) isochrones and $BVI$
photometry, and which incorporates
bright cluster members from Clari\'a (1982). There
were 98 correlations found within 6 arcseconds of the nominal X-ray
position and as discussed in Jeffries et al. (2006), where a similar
procedure was followed, these correlations are very
likely to be the genuine optical counterparts of the X-ray sources. 
The mean offset between the X-ray
and optical positions was 0.11 arcseconds in RA and 0.59 arcseconds in Dec.
As the optical catalogues have an absolute accuracy better than 0.2
arcseconds and an internal precision of about 0.05\,arcseconds, the X-ray
positions were corrected for these offsets. The remaining dispersion in
the offsets indicates an additional 1 arcsecond uncertainty (in
addition to the quoted astrometric uncertainty from the source searching
routines) in the X-ray positions, concurring with the current
astrometric calibration
assessment of the {\it XMM-Newton} science operations centre (Guainazzi 2010).

\subsection{Cross-correlation with the Irwin catalogue}

The purpose of this paper is to study the properties of fast-rotating 
M-dwarfs, so an investigation of the full X-ray source
population is deferred to a later paper. Here, we discuss
cross-correlations between the astrometrically corrected X-ray source
list and the catalogue of cool stars with known rotation periods in NGC
2547, given by Irwin et al. (2008). The Irwin et al. (2008) catalogue
contains precise positions tied to the 2MASS reference frame, rotation
periods and photometry in the $V$ and $I$ bands. We found significant
systematic differences (of up to 0.2 mag) between the photometry of
Irwin et al. (2008) and that found in Naylor et al. (2002) for stars
common to both.  As the accuracy of the Naylor et al. photometry has
support from an independent study by Lyra et al. (2006), we transformed
the Irwin photometry using the following best-fit relationships between
the Naylor et al. and Irwin et al. photometry:
\begin{equation}
V = 1.029\, V_{\rm Irwin} - 0.428 \ \ \ \ {\rm rms}=0.08\,{\rm mag},
\end{equation}
\begin{equation}
V-I = 1.177\, (V-I)_{\rm Irwin} - 0.305 \ \ \ {\rm rms}=0.10\,{\rm
    mag},
\end{equation}
where the calibrations transform the photometry onto the 
Johnson-Cousins calibration used by Naylor et al. and are valid
for $14<V<21$ and $1.2<(V-I)<3.4$. We take the rms deviation from these
relationships as the photometric uncertainty in subsequent analysis.

A maximum correlation radius between X-ray and optical sources of 5
arcseconds resulted in 68 correlations from the 97 Irwin et al. (2008)
objects that are within the EPIC field of view (there were no
correlations between 5 arcseconds and our nominal 6 arcsecond
acceptance threshold). The missing objects
were predominantly the optically faintest (see section 3.2). Experiments
involving offsetting the X-ray source positions by 30 arcseconds in
random directions suggest that fewer than 1 of these correlations would
be expected by chance. The details of the correlations along with count
rates, hardness ratios and correlation separations appear in Table~1
(available fully in electronic form only).

\subsection{Fluxes and coronal activity}

To assess magnetic activity levels, X-ray count rates were
converted into fluxes using a single conversion factor for each
instrument.  The use of a single conversion factor is necessitated because
few of the X-ray sources that correlate with the Irwin et
al. (2008) sample have sufficient counts ($>500$) to justify
fitting a complex spectral model. The median source is detected in the
EPIC-PN camera with about 150 counts and a signal-to-noise ratio of
only 10. Instead we designed a spectral model that
agrees with the mean hardness ratio (HR) in the EPIC-PN
camera, defined as
$(H-S)/(H+S)$ where $S$ is the 0.3--1.0\,keV count rate and $H$ is the
1.0--3.0\,keV count rate.

Our starting point was the analysis of an earlier, much shorter {\em
XMM-Newton} observation of NGC~2547 (Jeffries et al. 2006). This
showed that single-temperature thermal plasma models were
insufficiently complex to fit the spectra of active stars in
NGC~2547. A two-temperature ``{\sc mekal}'' model (Mewe, Kaastra \&
Leidahl 1995) provided a satisfactory
description, with $T_1 \simeq 0.6$\,keV, $T_2 \simeq 1.5$\,keV and an
emission measure ratio (hot/cool) of about 0.7, which was chosen to
approximately match the mean HR. This crude approximation
to the differential emission measure (DEM) of the coronal plasma
reasonably matches detailed work on the coronal DEM
of several nearby, rapidly-rotating low-mass stars, which show a
DEM maximum at around $10^{7}$\,K (Garc\'ia-Alvarez et al. 2008).  It was
also established in Jeffries et al. (2006) that the coronal plasma
was best fitted with a sub-solar metallicity ($Z \simeq 0.3$), which
seems to be a common feature for very active stars, including
fast-rotating K- and M-dwarfs (e.g. Briggs \& Pye 2003;
Garc\'ia-Alvarez et al. 2008).

Adopting the same model we have used the software package {\sc xspec}
and instrument response files appropriate for the EPIC-PN camera
at the time of the observations, to calculate a conversion factor from
0.3--3.0\,keV count rates to an {\em unabsorbed} 0.3--3.0\,keV flux. We
assumed an interstellar absorbing column density of neutral
hydrogen $N_H = 3\times 10^{20}$\,cm$^{-2}$ (see section 2.5). The derived flux
conversion factor is $1.68\times 10^{-12}$\,erg\,cm$^{-2}$ per count
and the derived HR is -0.42, which closely matches the mean HR in our
sample (see section~\ref{results}). Analogous conversion factors for the EPIC-MOS
cameras were derived by dividing the EPIC-PN conversion factor by the
weighted mean ratio of the observed MOS and PN count rates. This gave
conversion factors of $6.22\times10^{-12}$\,erg\,cm$^{-2}$ per count
for both of the EPIC-MOS cameras.

The unabsorbed 0.3--3\,keV X-ray flux for each detected Irwin et
al. (2008) star is found from a weighted average of fluxes from
each detector. Uncertainties in these fluxes arise from the
count rate errors, but we added a further 10 per cent systematic error
in quadrature to each detector count rate to account for uncertainties
in the instrument response and in the
the point spread function modelling in the {\sc edetect\_chain} task
(e.g. Saxton 2003; Guainazzi 2010). From the average fluxes the coronal
activity indicator $L_x/L_{\rm bol}$ was calculated, using the corrected $V$
magnitudes, an extinction $A_V=0.19$, a reddening $E(V-I)=0.077$ (see
Clari\'a 1982; Naylor et al. 2002) and
the relationship between intrinsic $V-I$ and bolometric correction
described by Naylor et al. (2002). For the red stars in our sample
with $V-I> 1.5$, these bolometric corrections are based on the empirical
measurements of Leggett et al. (1996). The bolometric corrections and 
$L_x/L_{\rm bol}$ values are reported in Table~1.

\begin{table*}
\caption{:\ X-ray detections of sources in the Irwin et al. (2008)
  catalogue of NGC~2547 members with rotation periods.  The Table is
  available electronically and contains 69 rows. Only two
  rows are shown here as a guide to form and content. Columns are as
  follows: (1) Identification from Irwin et al. (2008), (2-3) RA and
  Dec (J2000.0, from Irwin et al. 2008), (4-5) $V$ and $V-I$ magnitudes (modified from the
  Irwin et al. 2008 values -- see section 2.2), (6) rotation period
  (from Irwin et al. 2008), (7)
  $V$-band bolometric correction (see section 2.3), (8) $\log$
  bolometric luminosity (assuming a distance of 400\,pc), (9)
  $\log$ convective turnover time (see equation~3), (10) $\log$ Rossby Number, (11-12)
  stellar mass and radius (estimated from the Siess et al. 2000 models)
  (13) Keplerian co-rotation radius as a multiple of the stellar radius,
  (14-15) RA and Dec of the X-ray source, (16) the maximum log
  likelihood of the detection, (17)
  separation from the optical counterpart, (18-19) Total PN count rate
  (0.3--3\,keV), (20-21) PN count rate (0.3--1\,keV, the ``$S$'' band), (22-23) PN count
  rate (1--3\,keV, the ``$H$'' band), (24-25) PN hardness ratio
  (defined as $(H-S)/(H+S)$), (26-27) Total MOS1 count rate
  (0.3--3\,keV), (28-29) Total MOS2 count rate (0.3--3\,keV), (30-31) X-ray
  flux (0.3--3\,keV), (32-33) $\log$ X-ray to bolometric flux ratio,
  (34-35) cross identification with Naylor et al. (2002) where available.}
\begin{flushleft}
\begin{tabular}{lccc@{\hspace*{2mm}}ccc@{\hspace*{2mm}}c@{\hspace*{2mm}}c@{\hspace*{2mm}}c@{\hspace*{2mm}}c@{\hspace*{2mm}}c@{\hspace*{2mm}}c}
\hline
Name  & RA        &  Dec        & $V$ & $V-I$ & P & BC &$\log
L_{\rm bol}/L_{\odot}$& $\log \tau_c$ & $\log N_R$ & $M/M_{\odot}$ &
$R/R_{\odot}$ & $R_{\rm Kepler}/R$ \\
    &\multicolumn{2}{c}{(J2000)}&  &  & (d) & & & (d) & & & & \\
(1)  &               (2)       & (3)     & (4) & (5) &  (6) &    (7)
&(8) &  (9) &(10) & (11) &(12) &(13) \\
\hline
   N2547-1-5-417& 8 09 25.91& -49 09 58.4& 18.69& 2.80&  0.865&  -2.28& -1.40& 1.80& -1.86& 0.37& 0.47&   5.76\\
N2547-1-5-1123& 8 09 17.71& -49 08 34.5& 19.57& 3.16&  0.413&  -2.76& -1.56& 1.88& -2.26& 0.28& 0.42&   3.62\\
\hline
\end{tabular}
\vspace*{5mm}
\begin{tabular}{cccccccccccc}
\hline
 X-ray RA & X-ray Dec & ML & Sep & PN0 & PN0\_err & PN1 & PN1\_err &
 PN2 & PN2\_err & HR & HR\_err \\
\multicolumn{2}{c}{(deg, J2000.0)} & & (arcsec)& \multicolumn{2}{c}{(s$^{-1}$, 0.3--3\,keV)} &
 \multicolumn{2}{c}{(s$^{-1}$, 0.3--1\,keV)} &
 \multicolumn{2}{c}{(s$^{-1}$, 1--3\,keV)}& & \\
(14) & (15) & (16) & (17) & (18) & (19) & (20) & (21) & (22) & (23) & 
(24) & (25) \\
\hline
122.35782& -49.1664&  421.1& 0.62& 7.23e-03& 5.23e-04& 5.00e-03& 4.21e-04& 2.23e-03& 3.11e-04& -0.38&  0.07\\
122.32372& -49.1431&   36.4& 0.57& 1.84e-03& 3.76e-04& 1.33e-03& 2.99e-04& 5.04e-04& 2.28e-04& -0.45&  0.20\\
\hline
\end{tabular}
\vspace*{5mm}
\begin{tabular}{cccccc@{\hspace*{2mm}}c@{\hspace*{2mm}}cll}
\hline
MOS1 & MOS1\_err & MOS2 & MOS2\_err & Flux & Flux\_err & $\log
L_x/L_{\rm bol}$ & $\Delta \log L_x/L_{\rm bol}$ &
\multicolumn{2}{c}{Naylor ID}\\
\multicolumn{2}{c}{(s$^{-1}$, 0.3--3\,keV)} &
\multicolumn{2}{c}{(s$^{-1}$, 0.3--3\,keV)} &
\multicolumn{2}{c}{(erg\,cm\,s$^{-1}$, 0.3--3\,keV)} & & & &\\
(26) & (27) & (28) & (29) & (30) & (31) & (32) & (33) & (34) & (35) \\
\hline
1.82e-03 & 2.41e-04 & 1.80e-03& 2.32e-04& 9.59e-15&
 9.87e-16& -2.92&  0.08 & 14 & 1171\\
4.85e-04 &1.62e-04& 5.47e-04& 1.82e-04& 2.80e-15&
 6.05e-16& -3.30&  0.12 & 4  & 2832\\
\hline
\end{tabular}
\end{flushleft}
\label{xraydetect}
\end{table*}

\begin{table*}
\caption{The properties of stars from the Irwin et al. (2008) catalogue
  that have known rotation periods but were not detected within the
  {\it XMM-Newton} field. The Table is
  available electronically and contains 28 rows. Only two
  rows are shown here as a guide to form and content. The first 13
  columns contain the same properties as listed in
  Table~\ref{xraydetect}; following these, column 14 lists a flag
  denoting from which {\it XMM-Newton} instrument the X-ray
  count rate upper limit was derived: ``P'' for the PN, ``M1'' for
  MOS1, ``M2'' for MOS2 and ``M12'' for the average from MOS1 and MOS2
  (see text for details). Column 15 lists the 3-sigma count rate upper
  limit in that instrument, column 16 lists the corresponding flux
  upper limit (0.3--3\,keV) and column 17 lists the $\log L_x/L_{\rm
  bol}$ upper limit. Columns 18 and 19 list the cross identification
  with the Naylor et al. (2002) catalogue where available.}
\begin{flushleft}
\begin{tabular}{lccc@{\hspace*{2mm}}ccc@{\hspace*{2mm}}c@{\hspace*{2mm}}c@{\hspace*{2mm}}c@{\hspace*{2mm}}c@{\hspace*{2mm}}c@{\hspace*{2mm}}c}
\hline
Name  & RA        &  Dec        & $V$ & $V-I$ & P & BC &$\log
L_{\rm bol}/L_{\odot}$& $\log \tau_c$ & $\log N_R$ & $M/M_{\odot}$ &
$R/R_{\odot}$ & $R_{\rm Kepler}/R$ \\
    &\multicolumn{2}{c}{(J2000)}&  &  & (d) & & & (d) & & & & \\
(1)  &               (2)       & (3)     & (4) & (5) &  (6) &    (7)
&(8) &  (9) &(10) & (11) &(12) &(13) \\
\hline
 N2547-1-5-3742 & 8 09 25.73 & -49 03 15.7& 19.16& 2.92&  0.807&  -2.43& -1.53& 1.86& -1.96& 0.30& 0.43 &  5.64\\
  N2547-1-6-2487& 8 10 04.56 & -49 06 04.2& 14.26& 1.00&  5.488&  -0.27& -0.43& 1.31& -0.57& 0.88& 0.85 &  14.76\\
\hline
\end{tabular}
\vspace*{5mm}
\begin{tabular}{ccccll}
\hline
Instrument & Upper Limit & Flux & $\log L_x/L_{\rm bol}$ &
\multicolumn{2}{c}{Naylor ID}\\
           & ($s^{-1}$) & (erg\,cm\,s$^{-1}$, 0.3--3\,keV)& & & \\
(14) & (15) & (16) & (17) & (18) & (19)\\
\hline
P & $<$5.20e-03 & $<$8.73e-15 & $<-2.84$ & 4 & 1258\\
P & $<$1.16e-03 & $<$1.96e-15 & $<-4.59$ & 3 & 760\\
\hline
\end{tabular}
\end{flushleft}
\label{xraynodetect}
\end{table*}

\subsection{Flux upper limits}

There are 29 objects in the Irwin et al. (2008) catalogue within the
{\it XMM-Newton} field of view (for only a subset of the detectors in
some cases) but not found as sources by the {\sc edetect\_chain} task.
We inspected the X-ray images at the positions of these sources and
found one example (N2547-1-6-5108) which is a reasonably bright X-ray
source, that appears only in the MOS2 image, and which was missed by
the automated source searching. We evaluated the X-ray count rates for
this object using a 20 arcsecond radius aperture (see below) and a
local estimate of the background. This source has been added to
Table~1.  There were no X-ray sources apparent at the positions of the
other 28 objects and X-ray flux upper limits were derived as
follows. In the {\sc edetect\_chain} task, we generated images for each
detector consisting of smooth models of the X-ray background, to which
were added {\it models} of the significantly detected sources
calculated from their count rates and the model point spread function
used to detect and parameterise them. These images were ``noise-free''
estimates of the expected X-ray background for any given position.  The
total expected background was summed within circles of radius 20
arcseconds surrounding each of the undetected Irwin et al. (2008)
objects. The number of observed X-ray counts in those areas in the
original X-ray images was also summed, consisting of both source and
background counts.  A 3-sigma upper limit to the number of source
counts was calculated using the Bayesian approach formulated by Kraft,
Burrows \& Nousek (1991).  The choice of a 20 arcsec radius follows the
work of Carrera et al. (2007), who showed that ``aperture photometry''
using this radius gave count rates that closely matched those found in
{\sc edetect\_chain}. The upper limits to the X-ray count {\it rates}
were determined by dividing by the average exposure time within the
same circular area.

Using this technique we found that where objects were covered by both
PN and MOS data, that the PN data were at least a factor of two more
sensitive. Rather than attempt to combine the results from the three
instruments we have either (a) taken the upper limit from the PN, where
PN data are present (23 objects), (b) taken the the upper limit from one MOS
detector where an object was only covered by that detector (3 objects) or (c) taken
the average upper limit from both MOS detectors (2 objects) and divided by
$\sqrt{2}$, as in these cases the two upper limits were very similar.
The count rate upper limits were converted to upper limits in X-ray flux and
upper limits to $L_x/L_{\rm bol}$ using the procedures described in the
previous subsection. The details of the upper limit measurements are
presented separately in Table~2 (available in electronic form only).

\subsection{Uncertainties in X-ray fluxes}
The uncertainties in the $\log L_{\rm x}/L_{\rm bol}$ values
quoted in Table~1 incorporate the statistical count rate uncertainties
and the systematic instrument response uncertainties mentioned in
section~2.3. In addition we have included (in quadrature) uncertainties
in the bolometric flux due to the photometric uncertainties (or
variability) implied by equations~1 and~2, which amount to about $\pm
0.07$~dex. The final quoted uncertainties range from 0.08~dex to
0.21~dex, with a mean of 0.10~dex.

Other contributing uncertainties have also been considered.
The X-ray flux conversion factors are rather insensitive to the details of
the spectral model. For instance doubling the temperature of the hot
component increases the conversion factor by just 3 per cent and the HR
increases to -0.39; varying the metal abundance in the range $0.1<Z<1.0$
changes the conversion factor by only $\pm 5$ per cent; doubling the
ratio of hot-to-cool plasma emission measures  increases the conversion
factor by 3 per cent and increases the HR to -0.32. 
The value assumed for $N_H$ has a little more effect. The value of $N_H =
3\times10^{20}$\,cm$^{-2}$ is derived from the cluster reddening of
$E(B-V)=0.06\pm 0.02$ (Clari\'a 1982) and the relationship between
reddening and $N_H$ found by Bohlin, Savage \& Drake (1978), and could
be uncertain by factors of two. Altering $N_H$ from our assumed value
to either $1.5\times 10^{20}$\,cm$^{-2}$ or $6\times 10^{20}$\,cm$^{-2}$ would lead to
conversion factors about 5 per cent smaller or 10 per cent larger
respectively and HR would change to either -0.44 or -0.35 respectively.
All these uncertainties are small compared with those we have already
considered and they are neglected.

The final source of uncertainty is difficult to quantify, but is
probably dominant when considering a single epoch of X-ray data, namely
the coronal variability of active stars. Active, low-mass stars show
frequent X-ray flaring behaviour on timescales of minutes and hours,
resulting in an upward bias in an X-ray flux estimated, as here, over
the course of more than a day. The more active the low-mass star, the
more frequently it exhibits large X-ray flares (Audard et
al. 2000). Rather than try to correct for this, and as the comparison
samples in section~4 have not had any flare exclusion applied, we
continue to use the time-averaged X-ray flux to represent coronal
activity. An idea of the uncertainties can be gained by looking at
similar estimates from more than one epoch. For example, Gagn\'e,
Caillault \& Stauffer (1995) find that young Pleiades low-mass stars show differences of
more than a factor of two in their X-ray fluxes only 25 per cent of the
time on timescales of 1--10 years. Marino et al. (2003) find that the
median level of X-ray flux variability of K3-M dwarfs in the Pleiades
is about 0.2\,dex on timescales of months. Jeffries et al. (2006)
looked at the variability of G- to M-dwarfs in NGC~2547 itself on
timescales of 7~years, finding a median absolute deviation from equal
X-ray luminosity of about 0.1\,dex in G- and K dwarfs, with a hint that
the M-dwarfs are slightly more variable. Only about 20 per cent of
sources varied by more than a factor of two. Our conclusion is that we
should assume an additional uncertainty of about 0.1--0.2\,dex in our
single-epoch measurements, but caution that the distribution is
probably non-Gaussian in the sense that a sample will probably contain
a small number of objects that are upwardly biased by more than a
factor of two by large flares.

\section{Coronal activity in the M-dwarfs of NGC 2547}

\label{results}

\subsection{Hardness ratios}

\begin{figure*}
\centering
\begin{minipage}[t]{0.45\textwidth}
\includegraphics[width=80mm]{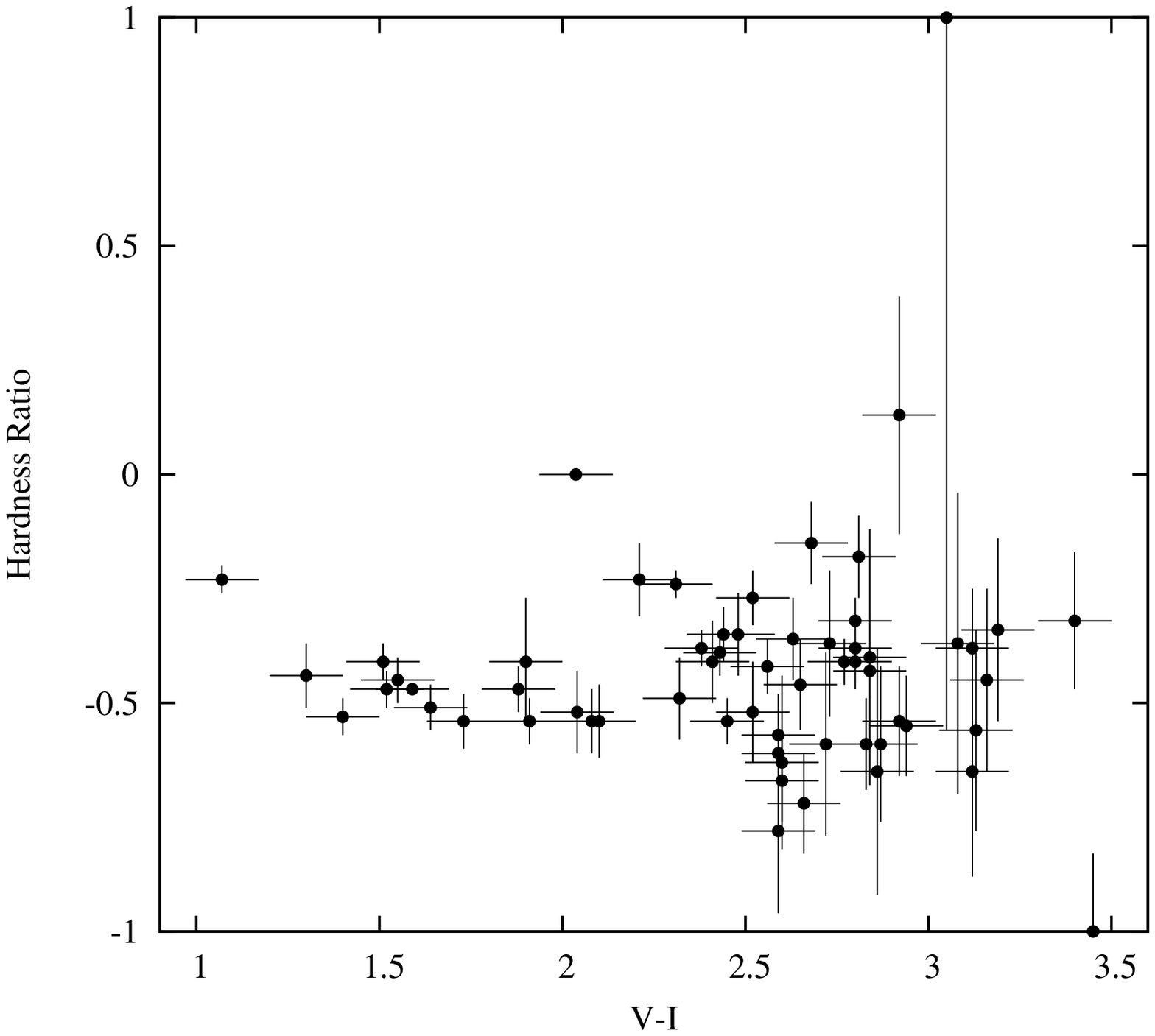}
\end{minipage}
\begin{minipage}[t]{0.45\textwidth}
\includegraphics[width=80mm]{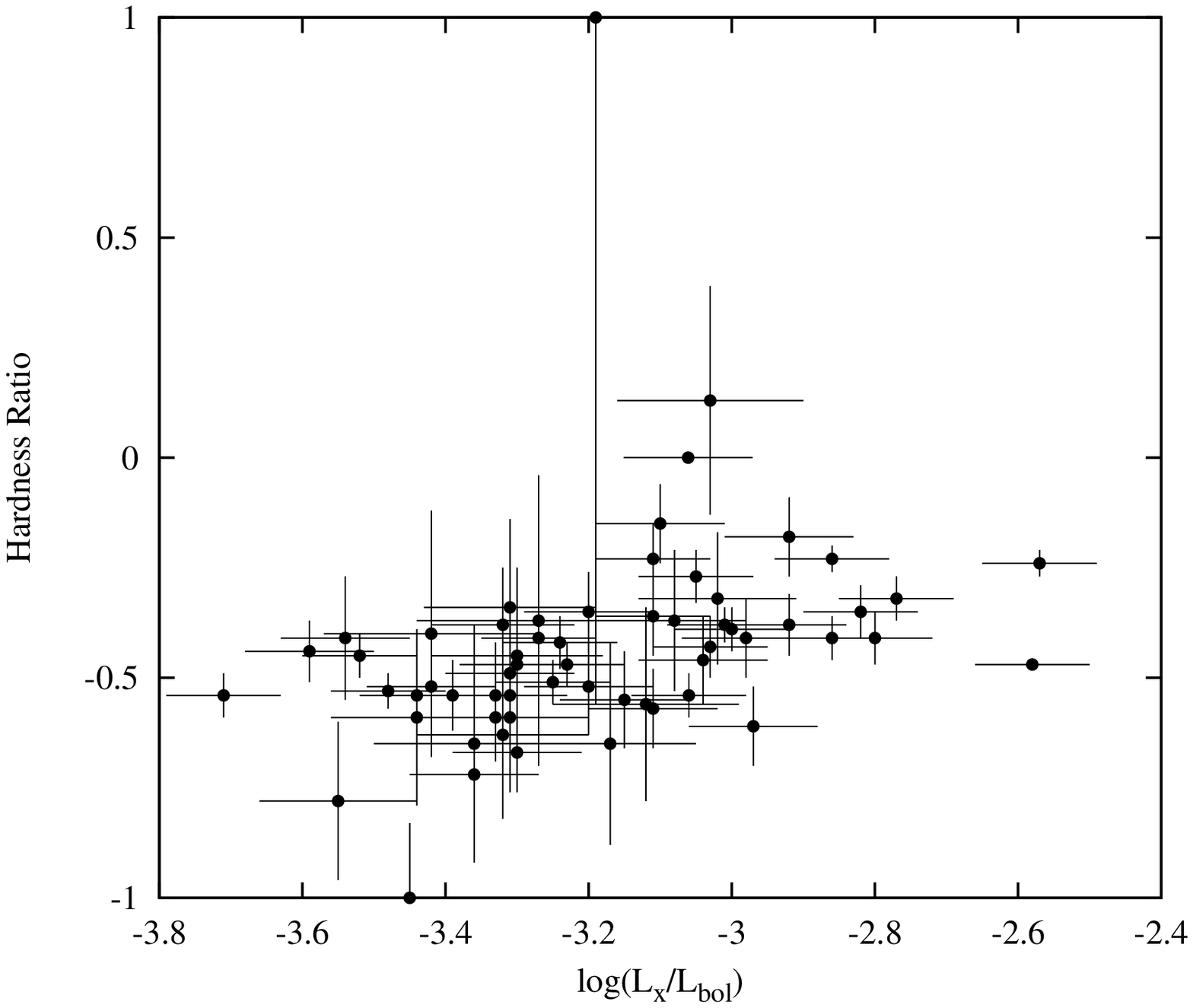}
\end{minipage}
\caption{Hardness ratios for the NGC~2547 sample 
  (defined as $(H-S)/(H+S)$ where $H$ is the
  1.0--3.0\,keV count rate and $S$ is the 
  0.3--1.0\,keV count rate in the EPIC-PN camera) as
  a function of (a) $V-I$ colour and (b) the X-ray to bolometric flux ratio.
}
\label{hardness}
\end{figure*}

In Figure~1 we show the dependence of the HR, measured by the PN
camera, on the intrinsic colour of the star and on the magnetic
activity level, expressed as $L_x/L_{\rm bol}$. There is no significant
dependence of HR on colour, and hence the approximation of a uniform
flux conversion factor with spectral type is shown to be
reasonable. The weighted mean HR is $-0.42\pm 0.02$, with a standard
deviation of 0.16. The standard deviation is about twice that expected
from the statistical errors, hinting at some genuine HR variation. Note
that this assumes that the HR uncertainties are symmetric and Gaussian,
even though the X-ray data are Poissonian in nature and HR is strictly
limited to $-1 \leq$\,HR\,$\leq 1$ (e.g. see Park et
al. 2006). The majority of sources are detected with
sufficiently large numbers of counts in both energy ranges and with
sufficiently precise count rates to make such an assumption
reasonable. The second plot shows there is some evidence that HR
increases slightly with activity level. A best-fit linear model is
HR$=-0.09(\pm 0.12) + 0.11(\pm 0.04) \log L_x/L_{\rm bol}$, but the
residuals suggest there may still be star-to-star variation at a given
activity level.  Note though that time variability of the X-ray
activity or HR values has not been considered (see section 2.5) and
this might plausibly account for some of this star-to-star variation.

A significant increase in average coronal temperatures (and hence HR)
with activity level is well established in field stars (e.g. Telleschi
et al. 2005). The relationship between HR and $L_x/L_{\rm bol}$ found
here and the size of the intrinsic scatter are indistinguishable from
those found for somewhat higher mass G- and K-stars in NGC~2547 by
Jeffries et al. (2006), as is the likelihood of an intrinsic
scatter. The mean HR agrees well (by design) with the HR predicted by
the spectral model used to calculate the X-ray fluxes. The probable
intrinsic variation in HR at a given $L_{x}/L_{\rm bol}$ corresponds
to variations of about a factor of two in the emission measure
ratio of the hot and cool plasmas in the two-temperature
coronal model. Such variations lead
to very small uncertainties in $L_{x}/L_{\rm bol}$ of order a few per
cent (see section 2.5) and so we have not attempted to correct for them.

\subsection{X-ray activity}

\begin{figure*}
\centering
\begin{minipage}[t]{0.45\textwidth}
\includegraphics[width=80mm]{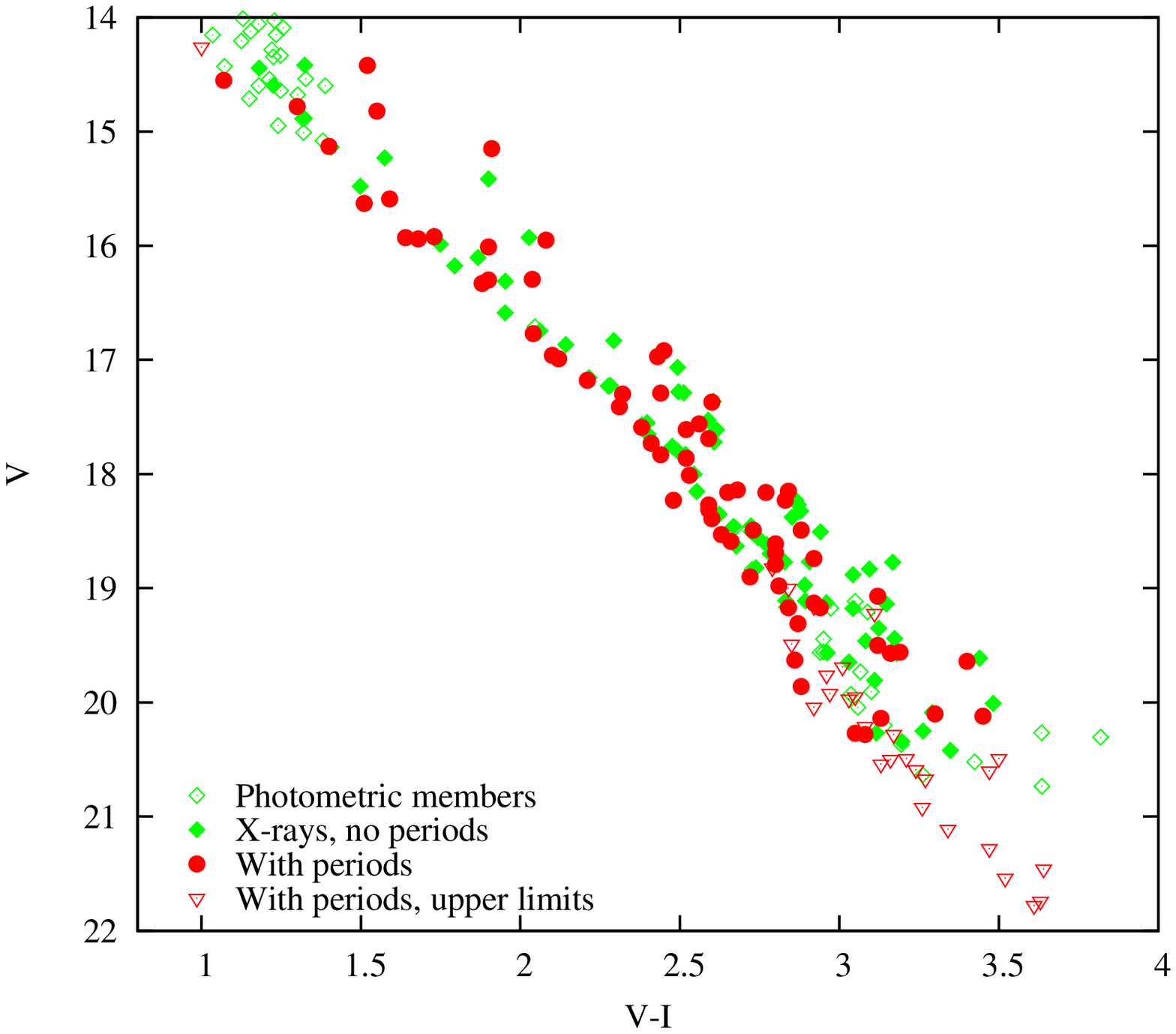}
\end{minipage}
\begin{minipage}[t]{0.45\textwidth}
\includegraphics[width=80mm]{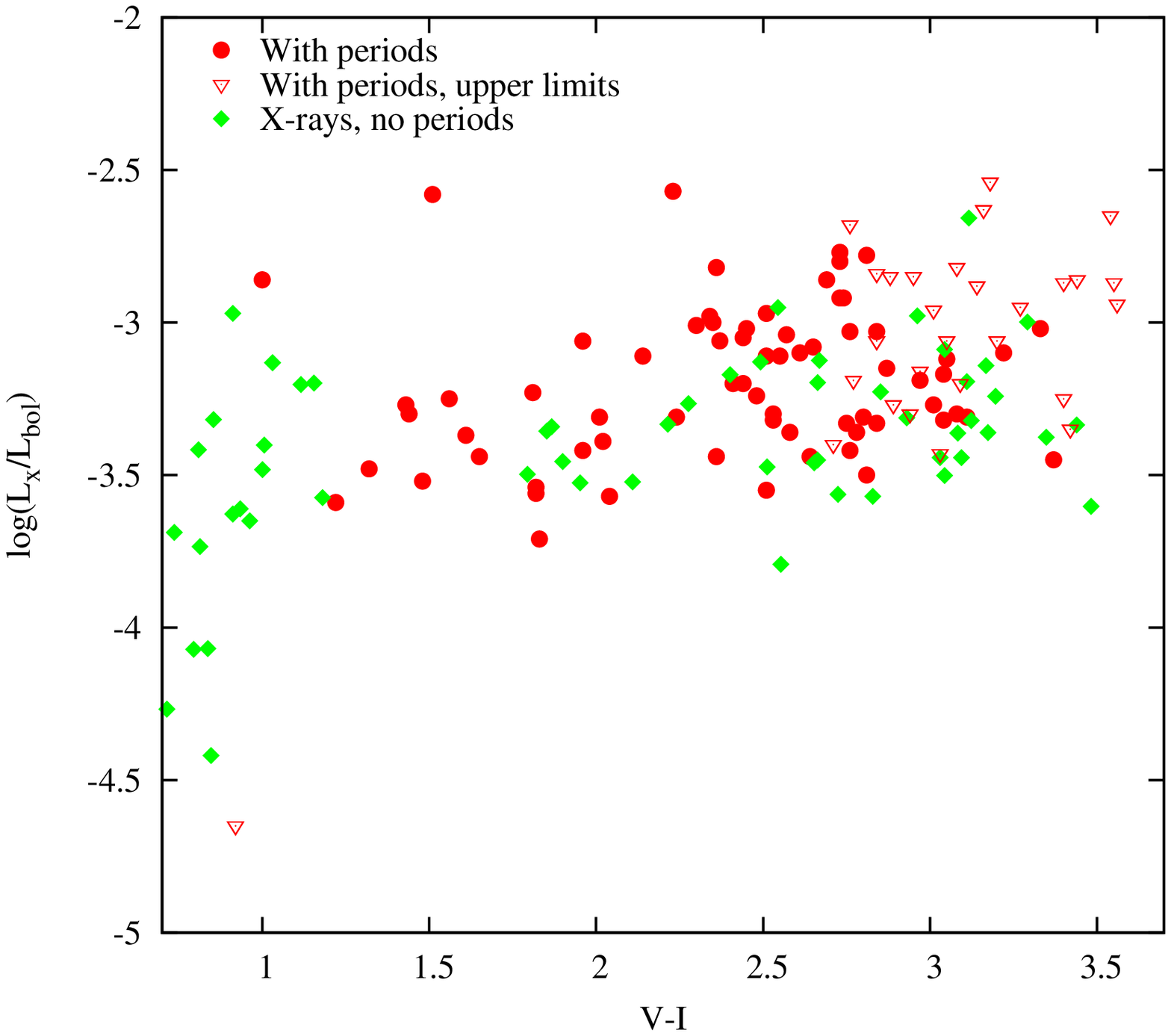}
\end{minipage}
\caption{(a) A colour-magnitude diagram in NGC~2547 showing: low-mass
  stars with rotation periods (from Irwin et al. 2008) that have
  significant X-ray detections (solid circles); stars with rotation
  periods, that are in the {\it XMM-Newton} field of view, but have
  only upper limits to their X-ray activity (open triangles);
  additional photometric candidate members of NGC~2547 (from Naylor et
  al. 2002) that are in the {\it XMM-Newton} field of view (open
  diamonds); and those additional photometric candidates that are
  detected in X-rays (filled diamonds). (b) X-ray activity, expressed
  as $\log L_{\rm x}/L_{\rm bol}$, as a function of colour. We compare
  the X-ray activity (and 3-sigma upper limits to X-ray activity) for
  stars in NGC~2547 with known rotation periods (from Irwin et
  al. 2008), to the X-ray activity of detected photometric members of
  NGC~2547 (from Naylor et al. 2002) that have no known rotation
  period.  Almost all photometric candidate members of NGC~2547 are
  X-ray detected for $V-I <2.8$ (see text), so we argue that the
  members with known rotation periods represent a reasonably unbiased
  sample with regard to X-ray activity and that the spread of activity
  at a given colour is less than an order of magnitude for $V-I
  >1.2$.
}
\label{lxlbol}
\end{figure*}

Figure~2a shows a $V$ versus $V-I$ colour magnitude diagram for the
stars of the Irwin et al. (2008) sample (with rotation periods) which
lie within the {\it XMM-Newton} field of view. As a comparison we show
photometric members of NGC~2547 (from Naylor et al. 2002) that also lie
within the X-ray field of view and indicate those which have an X-ray
counterpart among the 323 significant X-ray sources reported in
section~2.1. Figure~2a shows that the Irwin et al. (2008) sample is only
about 50 per cent complete. A significant fraction of photometic
cluster candidates at all colours have no detected rotation
periods. Most of these non-periodic candidates are X-ray detected and
very likely to be genuine NGC~2547 members (see Jeffries et al. 2006
for a detailed discussion). Only for stars with $V-I<1.5$ and $V-I>2.8$
are there significant numbers of candidates without X-ray
detections. For the former, these are likely to be contaminating
giants among the photometrically selected members, 
but the latter are more likely to be genuine cluster members
that become too faint for detection at very low luminosities (see
below), as indeed are many of the Irwin et al. (2008) objects at
similar colours.

Figure~2b plots X-ray activity, expressed as $\log L_{\rm x}/L_{\rm
  bol}$, versus colour. There is a gradually
  rising mean activity level as we move from K- through to M-dwarfs. The
  upper envelope is ill defined due to three very high points that seem
  well separated from the rest of the distribution. A time-series
  analysis of these three stars reveals that each was affected by a
  large flare during the course of the {\it XMM-Newton}
  observation. The lower envelope is well defined for $1< V-I <3$
  and probably represents a true floor to the level of X-ray activity
  in cool stars at the age of NGC~2547. Although our NGC~2547 sample is
  biased by having detected rotation periods, Irwin et al. (2008) argue
  that there is not a population of more slowly rotating low-mass stars
  and Fig.~2b also shows that photometrically selected members of
  NGC~2547 {\it without} rotation periods, and which also have a {\it
  XMM-Newton} counterpart, share a similar minimum (and median) level of
  X-ray activity as a function of colour.  Neither is this minimum
  level a product of the sensitivity of the X-ray observations, because
  almost all photometric candidates with $V-I < 2.8$ and which lie in
  the {\it XMM-Newton} field of view have been detected (see Fig.~2a).
 
The important point that emerges from these arguments is that the range
of X-ray activity levels in NGC~2547 at a given colour, is
quite small (less than an order of magnitude) for young K- and
M-dwarfs with $V-I<2.8$.  For cooler stars with $V-I>2.8$, the
X-ray observations become progressively incomplete and we can say very
little about the range of X-ray activity here.

\subsection{Activity versus rotation period and Rossby number}

\subsubsection{The dependence of activity on rotation period}

\begin{figure}
\centering
\includegraphics[width=80mm]{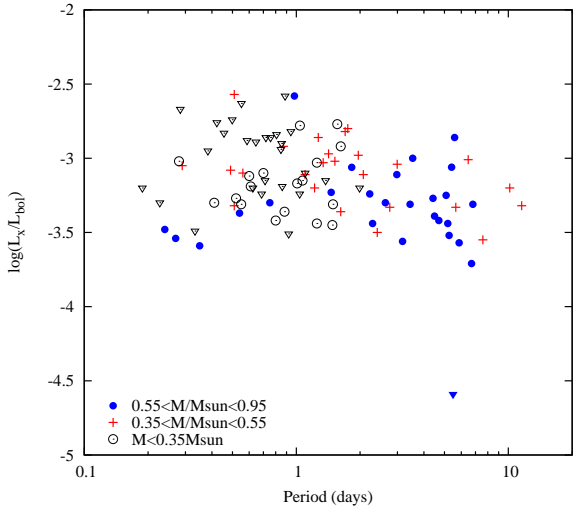}
\caption{X-ray activity as a function of rotation period for periodic
  objects in Irwin et al. (2008) that are in the {\it XMM-Newton} field
  of view. The sample has been split according to mass
  (estimated from the luminosity, see text); detections are denoted
  with symbols as shown in the plot and upper limits are shown with
  downward pointing triangles.
}
\label{lxlbolper}
\end{figure}

Figure~3 shows the dependence of activity ($L_{\rm x}/L_{\rm bol}$) on
rotation period, considering here only those stars in NGC~2547 with
rotation periods given by Irwin et al. (2008). For the purposes of
later discussion we have divided the stars up according to their
estimated masses: $0.55<M/M_{\odot}<0.95\,M_{\odot}$ (roughly
corresponding to K-stars, blue circles); $0.35<M/M_{\odot}<0.55$
(corresponding to M0-M3 stars, red crosses); and $M<0.35\,M_{\odot}$
(corresponding to stars cooler than M3, black open circles). The masses
were estimated from the luminosities of the stars and a 35\,Myr
isochrone from the evolutionary models of Siess, Dufour \& Forestini
(2000). The bolometric luminosities of our sample were calculated using
the corrected, intrinsic $V$ magnitudes, the bolometric corrections
described in section~2.3 and an assumed distance to NGC~2547 of 400\,pc
(e.g. Mayne \& Naylor 2008).

The choice of the mass division at $0.55\,M_{\odot}$ is to
isolate K-dwarfs from the cooler M-dwarfs and at $0.35\,M_{\odot}$ to
mark the approximate transition to a fully convective star (Siess et al. 2000).
More importantly, the latter division marks the lowest mass at which our sample can
be considered complete, in the sense that X-ray upper limits begin to
occur in stars of lower mass. The lowest luminosity stars in the sample
have masses of $\simeq 0.1\,M_{\odot}$.

Figure~3 shows that X-rays have been detected from K- and M-dwarfs with
periods ranging from 0.2--10\,days. Most of the upper limits occur for
stars with short periods. Given the possibility of a correlation
between X-ray activity and rotation this at first seems surprising.
The reason is that most of the lowest luminosity (and hence lowest
mass) objects in the Irwin
et al. (2008) catalogue have short rotation periods. As the the ratio $L_{\rm
x}/L_{\rm bol}$ is close-to-constant in this sample, this means that
many short period objects have low bolometric {\it and} X-ray
luminosities and are thus harder to detect.

There is a very shallow decline in X-ray activity
towards longer periods. As we will show when considering the Rossby
numbers for these stars, the reason that the longer period stars do
not show much lower X-ray activity is that most of them still have
small Rossby numbers and are in the regime where saturated activity
levels are expected.

There is some evidence that the very shortest period stars might have
lower activity corresponding to ``super-saturation''.  This is
primarily based on a group of three very low-mass stars with $P<0.4$\,d
and only upper limits to their X-ray activity and a similar group of
three K-stars ($M>0.55\,M_{\odot}$) where $L_x/L_{\rm bol} \simeq 10^{-3.5}$.

Any correlations, trends or threshold periods may be confused or blurred
by the inclusion of a range of spectral types/convection zone depths
within the sample plotted in Fig.~3. It was exactly this issue which
led previous authors to consider the use of the Rossby number
as a proxy for magnetic dynamo
efficiency (e.g. Noyes et al. 1984; Dobson \& Radick 1989; Pizzolato et
al. 2003).

\subsubsection{Rossby numbers}

\begin{figure}
\centering
\includegraphics[width=80mm]{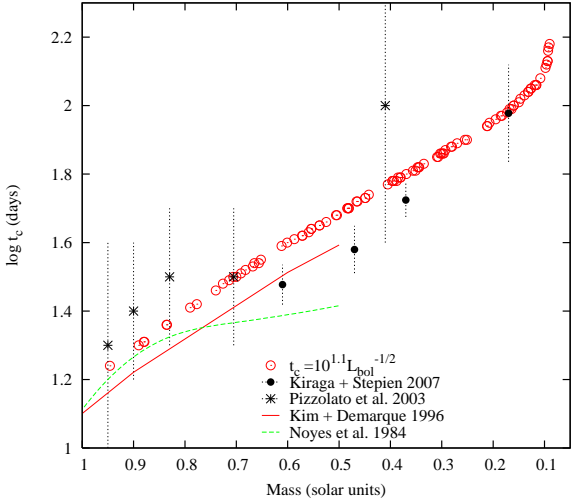}
\caption{Convective turnover time as a function of stellar mass. The
  open circles show the turnover times adopted for the NGC~2547 sample,
  calculated according to $\tau_c \propto L_{\rm bol}^{-1/2}$ (see
  text), compared with empirical calibrations from Noyes et al. (1984),
  Pizzolato et
  al. (2003) and Kiraga \& St{\c e}pie\'n (2007) and a scaled theoretical 200\,Myr
  isochrone from Kim \& Demarque (1996). The masses for the NGC~2547
  stars were estimated from their luminosities (see text).
}
\label{tauc}
\end{figure}

The use of Rossby number raises a problem when dealing with
M-dwarfs. The widely used semi-empirical formula of Noyes et al. (1984)
predicts convective turnover time, $\log \tau_c$, as a function of
$B-V$ colour. This relationship is poorly defined for $B-V>1$ and has
no constraining data in the M-dwarf regime. Theoretically, little work
has been done on turnover times at very low masses.  Gilliland (1986)
calculated that the turnover time at the base of the convection zone
increases with decreasing $T_{\rm eff}$ and mass along the main
sequence.  These models are limited to stars $\geq 0.5\,M_{\sun}$,
corresponding to $T_{\rm eff} \geq 3500$\,K on the main sequence.
$\log \tau_c$ is roughly linearly dependent on $\log T_{\rm eff}$ and
increases from about 1.2 (when $\tau_c$ is expressed in days) at
$T_{\rm eff}=5800$\,K to about 1.85 at 3500\,K. Similar calculations,
with similar results (except for arbitrary scaling factors) have been
presented more recently by Kim \& Demarque (1996), Ventura et
al. (1998) and Landin, Mendes \& Vaz (2010). Ventura et al. (1998) attempted to extend the calculation into
the fully convective region, predicting that the convective turnover
time would continue to increase.  A complication here is that the
M-dwarfs of NGC~2547 are not on the main sequence. Gilliland
(1986), Kim \& Demarque (1996) and Landin et al. (2010) 
predict that $\tau_c$ is about 50 per
cent larger for stars of $0.5\,M_{\odot}$ at an age of $\sim 30$\,Myr on
the pre-main-sequence (PMS).

An alternative approach has been to empirically determine $\tau_c$ by
demanding that activity indicators (chromospheric or coronal) satisfy a
single scaling law with Rossby number, irrespective of stellar mass
(e.g. Noyes et al. 1984; St{\c e}pie\'n 1994). The most recent work has
focused on $L_{\rm x}/L_{\rm bol}$ as an activity indicator. Using 
slow- and fast-rotating stars, Pizzolato et al. (2003) 
showed that $\tau_c$ needs to increase rapidly with decreasing
mass in order to simultaneously explain $L_{x}/L_{\rm
bol}$ in G-, K- and M-dwarfs, and they find $\log \tau_c > 2$ for
$M<0.5\,M_{\odot}$. Similar work by Kiraga \& St{\c e}pie\'n (2007)
concentrated on slowly rotating M-dwarfs finding their empirical $\log
\tau_c$ increased from 1.48 at $M\simeq 0.6\,M_{\odot}$ to 1.98 at $M
\simeq 0.2\,M_{\odot}$. An interesting insight was
provided by Pizzolato et al. (2003), who noted that the mass dependence
of the turnover time is closely reproduced by assuming 
that $\tau_c \propto L_{\rm bol}^{-1/2}$.
In what follows we adopt this scaling and hence
a Rossby number given by
\begin{equation}
\log N_R = \log P - 1.1 + 0.5 \log L_{\rm bol}/L_{\odot}\, .
\end{equation}
The turnover time has been anchored such that $\log \tau_c = 1.1$
for a solar luminosity star, following the convention adopted by Noyes et
al. (1984) and Pizzolato et al. (2003). 

In Fig.~\ref{tauc} we compare the $\tau_c$ values calculated for our
sample as a function of mass, with the theoretical and empirical
relationships discussed above. Overall, our $\tau_c$ values lie a little below
(but not significantly) the Pizzolato et al. (2003)
calibration, and marginally above the values determined by Kiraga \& St{\c e}pie\'n
(2007) and a theoretical 200\,Myr isochrone from Kim \& Demarque
(1996).\footnote{The $\tau_c$ values from Kim \& Demarque (1996) were
multiplied by 0.31 to anchor them at $\log \tau_c = 1.1$ for a solar
mass star.} From the discussion by these latter authors and their
Fig.~3, it is clear that any discrepancy between theory and
data could be explained by the younger age of NGC~2547. Stars with
$M<1.0\,M_{\odot}$ are still approaching the main sequence at
30--40\,Myr, their $\tau_c$ values are still falling and one would
expect them to have larger $\tau_c$ than for a 200\,Myr isochrone or
indeed M-dwarf field stars.  Unfortunately, Kim \& Demarque (1996) did
not provide $\tau_c$ isochrones at these younger ages.
The semi-empirical relationship between $\tau_c$ and mass proposed by
Noyes et al. (1984) is also shown. As discussed above, this has little
or no constraining data below about $0.7\,M_{\odot}$ and our adopted
turnover times in this regime are much larger.

We conclude that our adopted $\tau_c$ values and hence Rossby numbers
are quite consistent with previous work, although systematic
uncertainties at the level of $\sim 0.2$\,dex must be present when
comparing stars at $M<0.5\,M_{\odot}$ with solar-mass stars. Tables
1~and~2 include our calculated convective turnover times and Rossby
numbers.

\subsubsection{The dependence of activity on Rossby number}

\begin{figure}
\centering
\includegraphics[width=80mm]{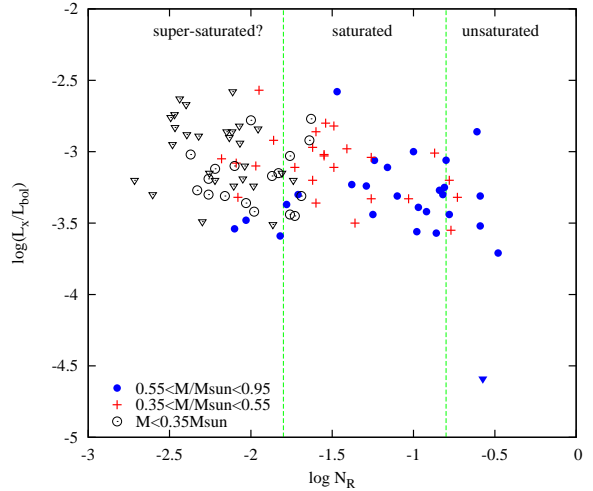}
\caption{X-ray activity as a function of Rossby number. The sample is
  divided by mass and shown with symbols as in Fig.~\ref{lxlbolper}. The dashed lines
  indicate divisions between unsaturated, saturated and
  super-saturated coronal activity defined in the literature for G- and K-stars (see section
  3.3.3).
}
\label{lxlbolrossby}
\end{figure}

Figure~\ref{lxlbolrossby} shows the dependence of $L_{\rm x}/L_{\rm
bol}$ on Rossby number. Dashed loci indicate the approximate divisions
between the unsaturated, saturated and
super-saturated regimes found from data for G- and K-stars (see Randich
1998; Pizzolato et al. 2003 and section~4). There is some evidence that
the K-stars in our sample ($M>0.55\,M_{\odot}$, blue circles) are following the
pattern seen in other young clusters. Most of the K-dwarfs appear to
have a saturated level of magnetic activity, although we do not clearly
see any fall in activity levels at the largest Rossby numbers in our
sample.  There are three K-dwarfs with $L_{\rm x}/L_{\rm bol}
\simeq 10^{-3.5}$ at $\log N_R <-1.8$, which may be examples of
super-saturated stars, and whilst there are also some with similar
activity levels at higher Rossby numbers it is clear that the K-dwarfs
with the lowest Rossby numbers are less active than their equivalents
in the M-dwarf subsample. The 5 K-dwarfs with $\log N_R < -1.7$ have a
mean $\log L_{\rm x}/L_{\rm bol}= -3.46 \pm 0.11$ (standard deviation),
while the 7 M-dwarfs (with $M>0.35\,M_{\odot}$) at similar Rossby numbers have a mean $\log
L_{\rm x}/L_{\rm bol}= -3.02 \pm 0.21$.

In the M-dwarf data for NGC~2547 ($M<0.55\,M_{\odot}$, red crosses and
open circles) there is no
decline in activity at large Rossby numbers.
This is probably due to a lack of targets
with long enough periods to populate
the unsaturated regime ($\log N_R > -0.8$, Stauffer et al. 1997).
However, there is no doubt that an unsaturated coronal activity regime
does exist for field M-dwarfs with long rotation periods (see Kiraga \&
St{\c e}pie\'n 2007 and section~4). 

The lack of obvious 
super-saturation among the M-dwarfs is more
significant. The phenomenon is claimed to begin in G- and
K-dwarfs with $\log N_R\leq -1.8$ (Patten \& Simon 1996; Stauffer et
al. 1997; Randich 1998) and there are plenty of stars in our sample
with much smaller Rossby numbers than this, mostly among the lowest
mass M-dwarfs.  A complication here is that
the cited papers used relationships in Noyes et al. (1984) to calculate
$\tau_c$. However, Fig.~\ref{tauc} shows that Rossby numbers estimated
for G- and K-dwarfs ($M>0.55\,M_{\odot})$ using the Noyes et al. (1984)
formulae would be $\leq 0.2$\,dex larger than those calculated in
this paper for a star of similar mass, so the conclusion is robust.

It is possible that super-saturation begins in the M-dwarfs of NGC~2547
at very low Rossby numbers, say $\log N_R <-2.3$, but this is 
only indicated by three upper limits.  James et al. (2000) also claimed
tentative evidence of super-saturation in their collected sample of
field and cluster M-dwarfs. Their claim was based on two stars;
VXR47 and VXR60b in IC~2391, which have $(V-I)_0=1.7$ and 2.06
respectively and according to their luminosities would both be
classed as late K-dwarfs  ($M>0.55\,M_{\odot}$) with $\log N_R \simeq
-2$ in our classification scheme.

Our conclusion from the NGC~2547 sample is that if super-saturation
does occur in M-dwarfs it occurs at Rossby numbers that become (in the
lowest mass stars) at least a
factor of three lower than in super-saturated G- and K-dwarfs.

\section{Combination with other datasets}

The NGC~2547 data suggest that super-saturation does not occur at a
single critical Rossby number irrespective of spectral type (see
Fig.~5). Instead it seems more plausible that super-saturation may
occur below some critical period ($\leq 0.4$~d, see Fig.~3). This
interpretation is hampered by relatively few targets with very short
rotation periods and also few targets of higher mass, where the very
different convective turnover times would result in quite different
Rossby numbers at the same rotation period. To increase our statistics
and to make a better comparison with a larger sample of higher mass
stars, we combined our dataset with published data for stars with known
rotation periods in young open clusters and the field. The main sources
of the comparison data are catalogues of X-ray activity in field
M-dwarfs by Kiraga \& St{\c e}pie\'n (2007) and in field and cluster
dwarfs by Pizzolato et al. (2003). We have also used the
the recent catalogue of rotation periods in Hartman et al. (2010, see
below) to add many new stars from the young Pleiades cluster to the
rotation-activity relationship. This large sample covers a range of
ages:  NGC~2547 at 35\,Myr (Jeffries \& Oliveira 2005); IC~2391 and
IC~2602 at $\simeq 50$\,Myr (Barrado y Navascues, Stauffer \&
Jayawardhana 2004); Alpha Per at $\simeq 90$\,Myr (Stauffer et al. 1999);
Pleiades at $\simeq 125$\,Myr (Stauffer, Schultz \& Kirkpatrick 1998);
Hyades at $\simeq 625$\,Myr (Perryman et al. 1998); and field stars that
have ages from tens of Myr to many Gyr.

There are important issues to address regarding the completeness of
comparison samples. For NGC~2547, we started with stars of known
rotation period and determined their X-ray activity level or upper
limits to their X-ray activity if they could not be detected. Almost
all stars were detected down to about $0.35\,M_{\odot}$ (spectral type
M3) and there were a mixture of detections and upper limits at lower
masses.  The samples of stars with known X-ray activity and rotation
period found in the literature have been generated in a different
way. Generally, field star samples have been compiled by matching
objects with known rotation periods to X-ray sources in the {\it ROSAT}
all-sky survey (RASS, Voges et al. 1999). In clusters, the data are
mostly from {\it ROSAT} pointed observations, from which detected X-ray
sources were matched with known cluster members. The problem is,
especially when searching for evidence of super-saturation, that we
must be sure that the X-ray observations were sensitive enough to have
detected examples of super-saturated stars (say with $L_x/L_{\rm bol}
\leq 10^{-3.5}$). This difficulty is exacerbated for active M-dwarfs,
because they frequently flare, so there is a possibility that what has
been reported in the literature is an X-ray bright tail, disguising a
hidden population of undetected, super-saturated M-dwarfs.

\begin{figure*}
\centering
\begin{minipage}[t]{0.45\textwidth}
\includegraphics[width=80mm]{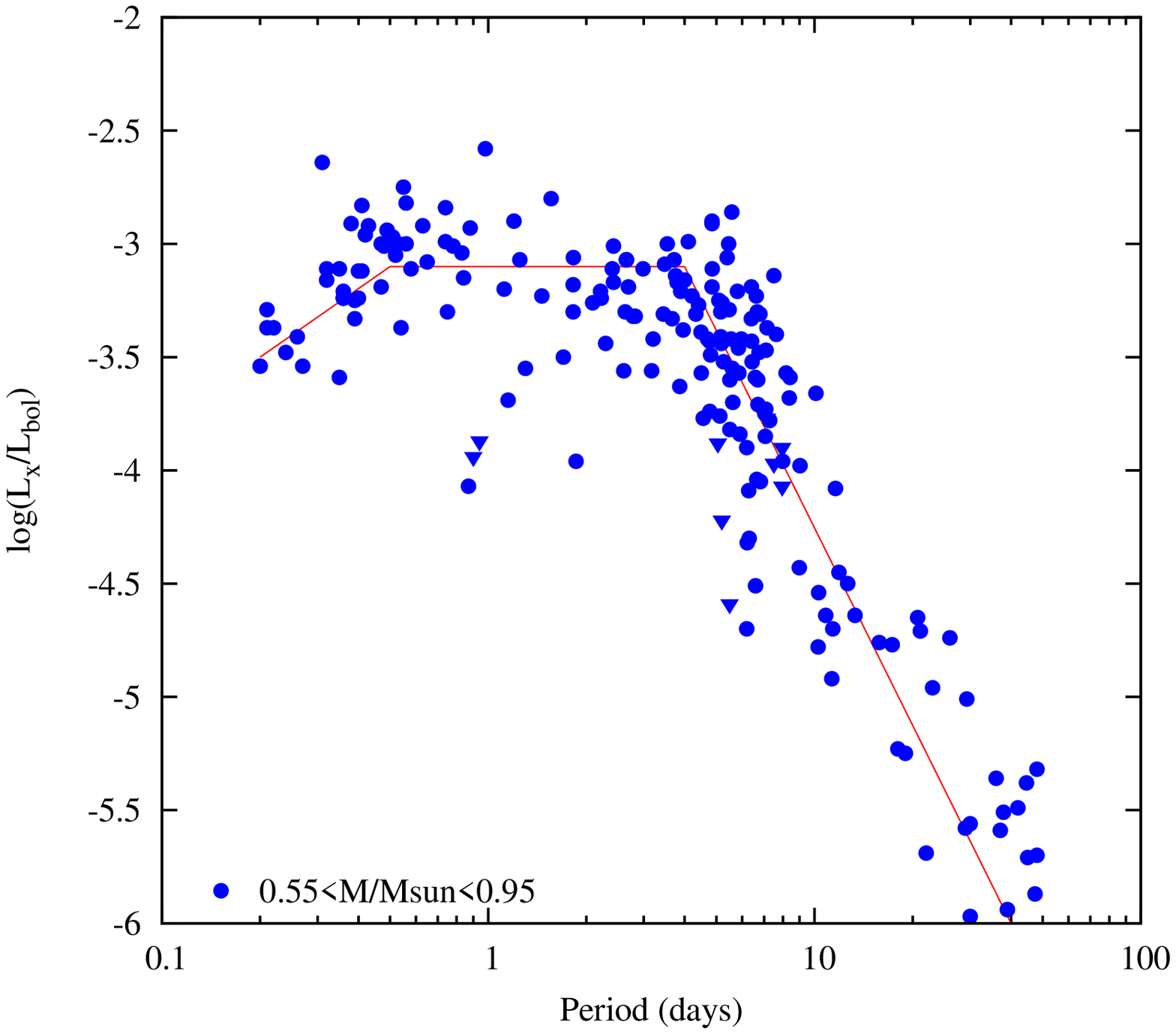}
\includegraphics[width=80mm]{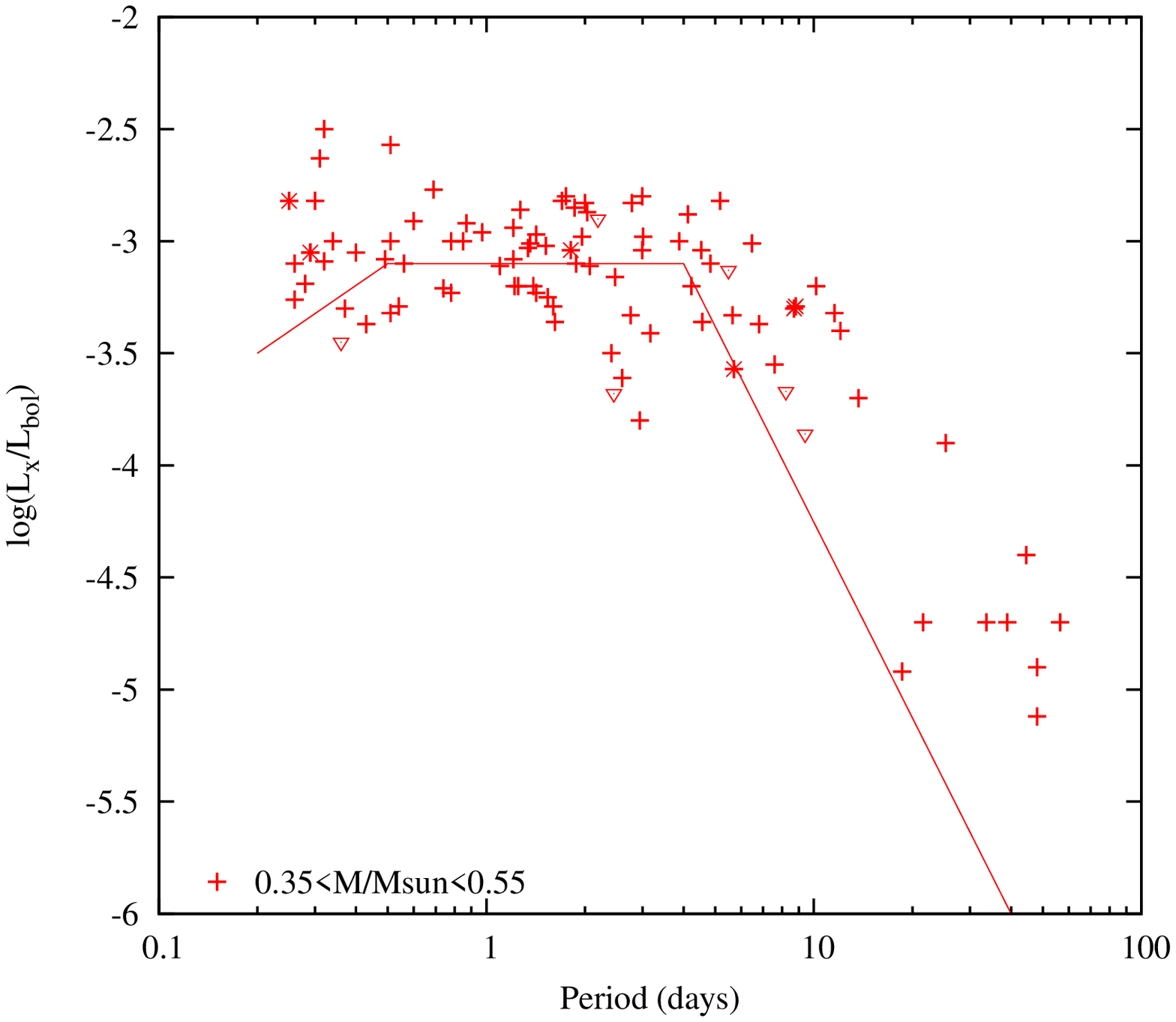}
\includegraphics[width=80mm]{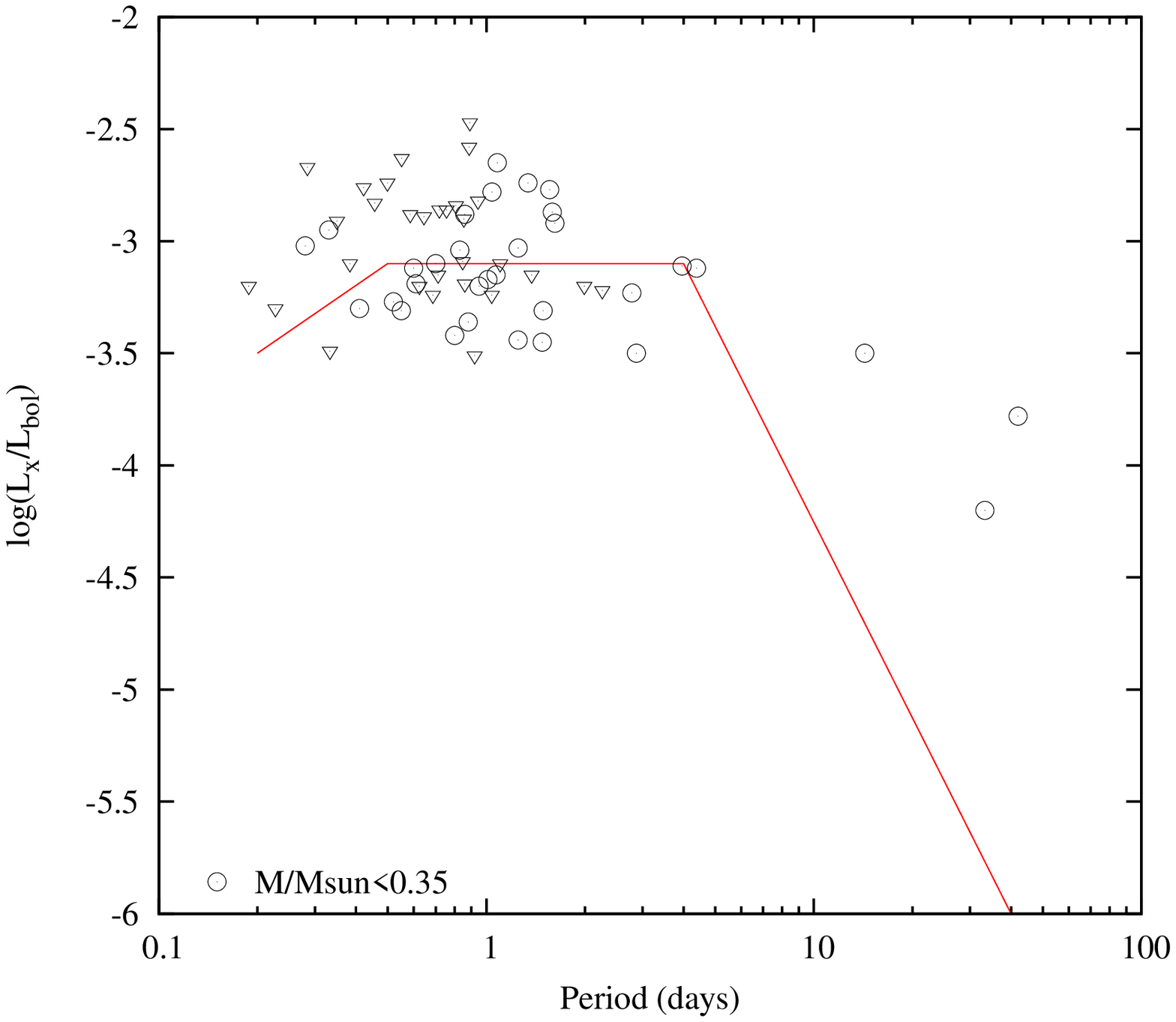}
\end{minipage}
\begin{minipage}[t]{0.45\textwidth}
\includegraphics[width=80mm]{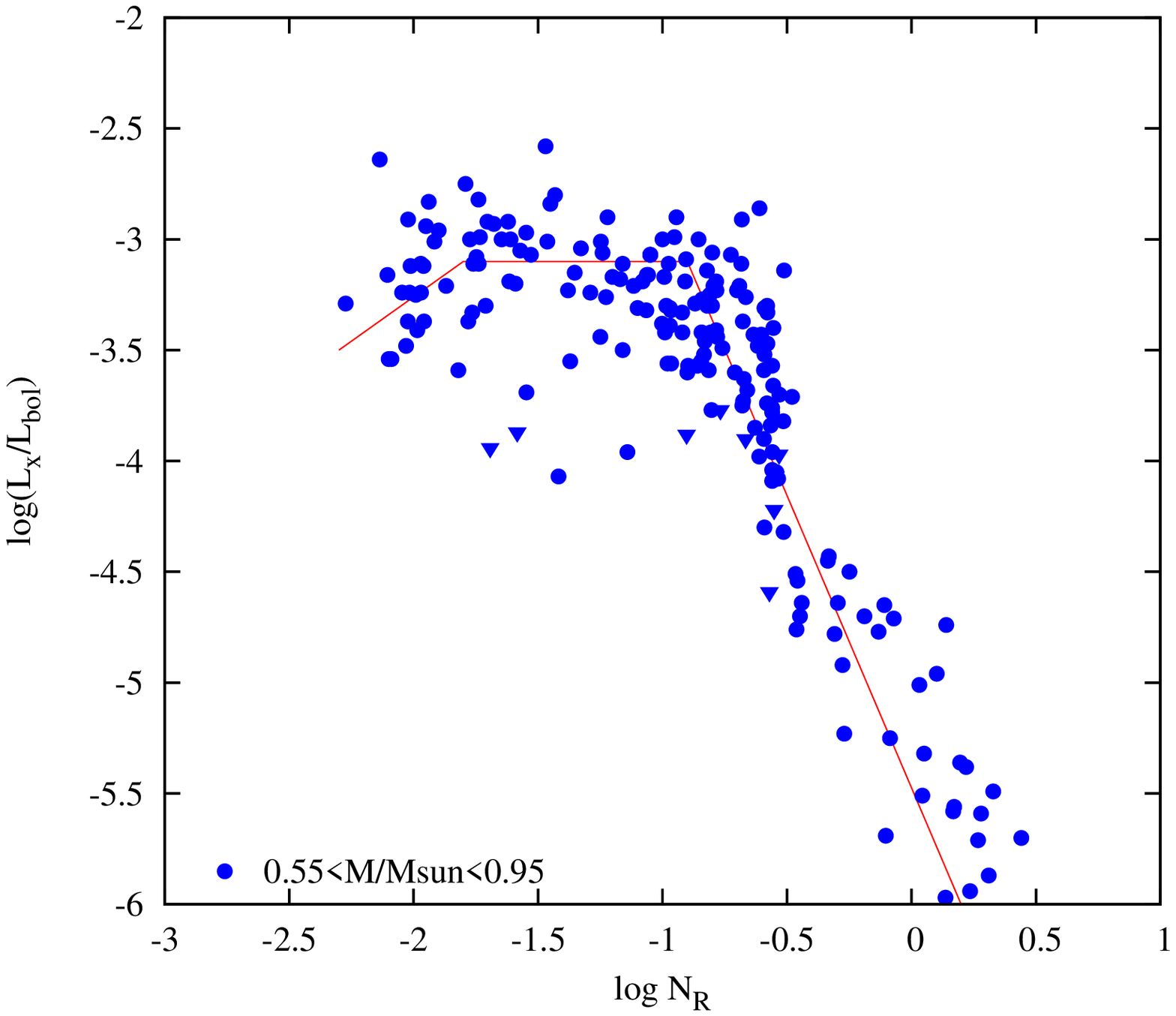}
\includegraphics[width=80mm]{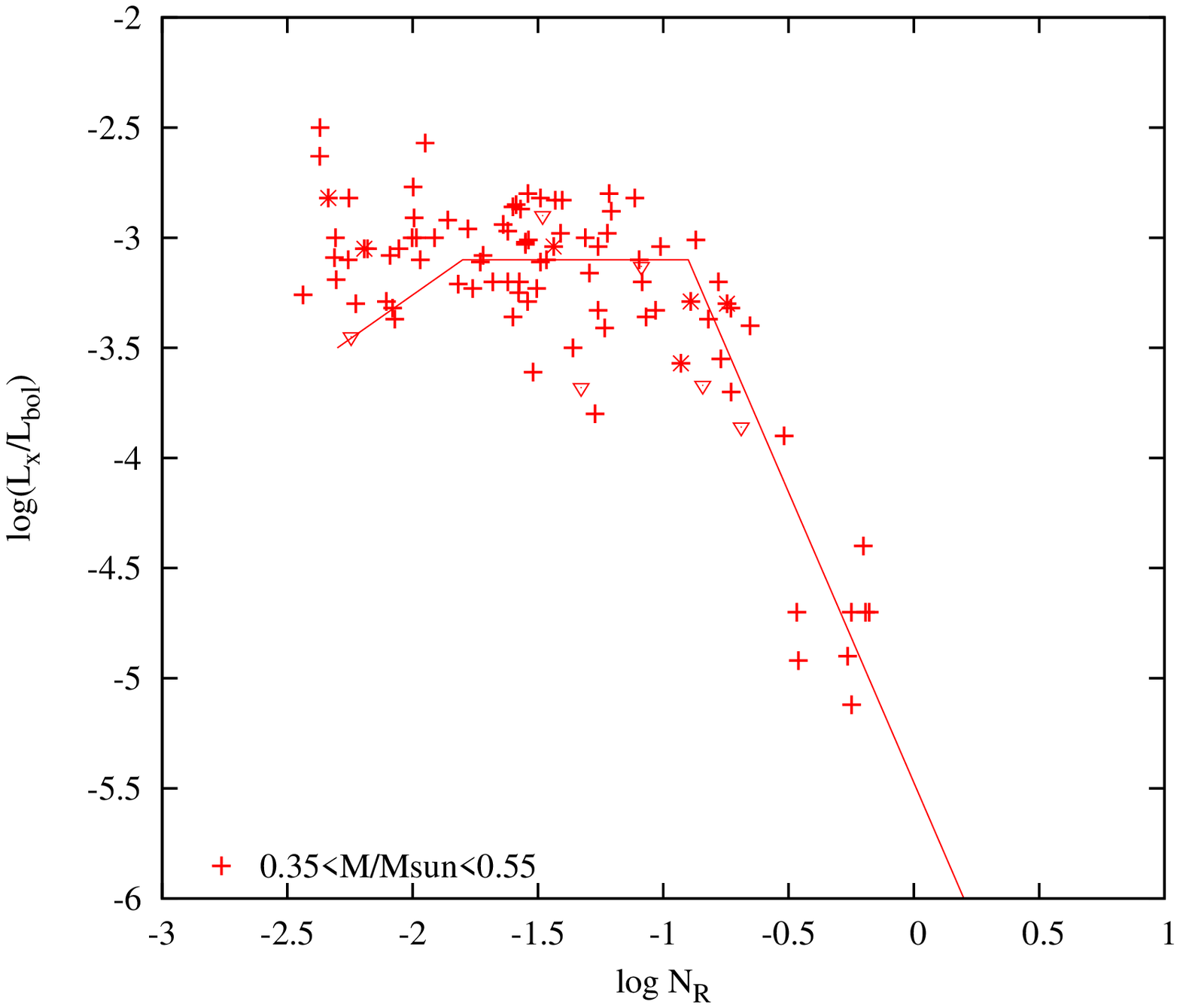}
\includegraphics[width=80mm]{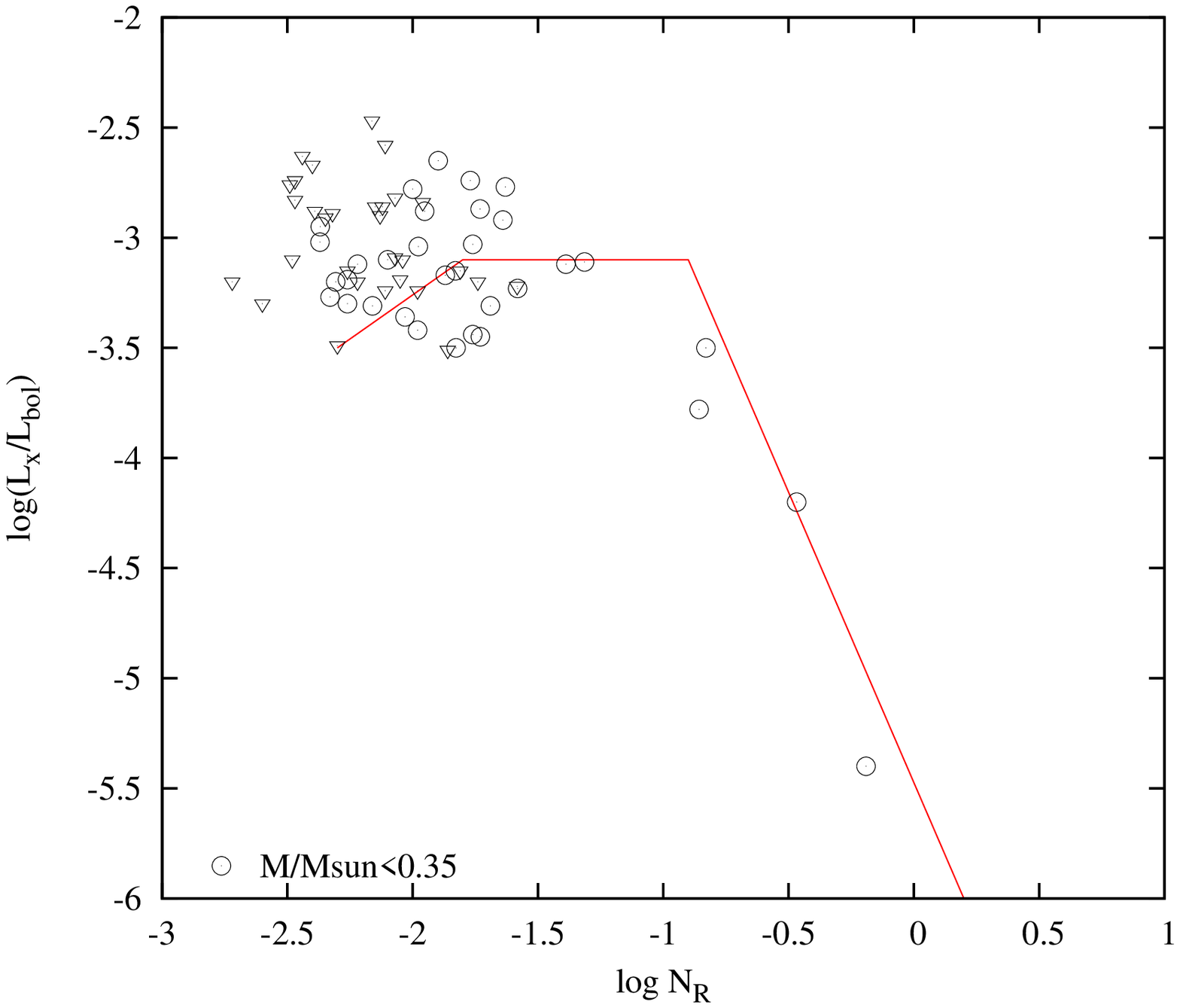}
\end{minipage}
\caption{X-ray activity as a function of rotation period (left column)
  or as a function of Rossby number (right column) for
  low-mass stars in NGC~2547 and for literature compilations of cluster
  and field dwarfs as explained in the text.
  The plots are separated according to the estimated masses of the stars
  (top: K-stars with $0.55<M/M_{\odot}<0.95$; middle: M-stars with
  $0.35<M/M_{\odot}<0.55$; bottom: M-stars with $M<0.35\,M_{\odot}$,
  the symbols are those used in Fig.~3).
  The loci in each plot are by-eye trends identified in the K-stars and
  then repeated in the subsequent plots to highlight differences with
  stellar mass. Cluster M-dwarfs from Pizzolato et al. (2003) may be
  subject to an upward bias (see text) and are marked with asterisks.
}
\label{comblxlbol}
\end{figure*}

For field stars at a given $V$ magnitude, and for a given X-ray flux
detection threshold, the corresponding $L_x/L_{\rm bol}$
detection threshold becomes smaller in cooler stars because the magnitude of
the $V$-band bolometric correction increases. 
The catalogue of M-dwarfs with rotation periods compiled by Kiraga \&
St{\c e}pie\'n (2007) was taken from optical samples with $V<12.5$
and correlated with {\it ROSAT} data, mainly from the RASS. The flux
sensitivity of RASS varies with ecliptic latitude, but the minimum
detectable count rate over most of the sky is $\simeq 0.015$ counts per
second. For a typical coronal spectrum, and neglecting interstellar
absorption for nearby stars, this equates to a flux
sensitivity of $10^{-13}$ erg\,cm$^{-2}$\,s$^{-1}$ (e.g. H\"unsch et
al. 1999).  If we consider an M0 dwarf with $V=12.5$, this flux
detection limit corresponds to $L_{x}/L_{\rm bol} \simeq 1.3 \times
10^{-4}$ and is even lower for cooler M-dwarfs. Hence the Kiraga
\& St{\c e}pie\'n field M-dwarf sample was easily capable of
identifying super-saturation. Similarly, the faintest field M-dwarfs in
the Pizzolato et al. (2003) compilation have $V \sim 11$ and would easily be
detected at super-saturated activity levels in the RASS. There are thus
no completeness problems for the comparison samples of field dwarfs.

In a cluster, all stars are at the same distance, so a given X-ray
flux detection limit corresponds to a $L_{x}$ threshold, and cooler
stars with smaller $L_{\rm bol}$ are harder to detect at a
given $L_x/L_{\rm bol}$. In most clusters observed by
{\it ROSAT}, and included in the Pizzolato et al. compilation, the
sensitivity was sufficient to detect all G- and K-dwarfs, but only
a fraction of M-dwarf members were detected (e.g. Stauffer et
al. 1994; Micela et al. 1999 - the Pleiades; Randich et al. 1995 -
IC~2602; Randich et al. 1996 - Alpha Per). Furthermore in some clusters
(e.g. IC 2391 -- Patten \& Simon 1996), members were {\it
identified} on the basis of their X-ray detection and their rotation
periods were determined afterwards. In these circumstances it is
likely that the M-dwarfs identified in the cluster observations will
have mean activity levels biased upwards by selection effects and there
would be little chance of identifying super-saturation even if it were
present. The cluster M-dwarfs from Pizzolato et al. (2003) are
therefore clearly identified in what follows.

In the case of the Pleiades, these
difficulties were circumvented by replacing those Pleiades stars in Pizzolato et
al. (2003) with a sample constructed by cross-correlating the
Hartman et al. (2010) catalogue of rotation periods for G-, K- and
M-dwarfs with the {\it ROSAT} source catalogues
(including upper limits) of Micela et al. (1999) and Stauffer et
al. (1994) (in that order of precedence where sources were detected in
both). In this way we constructed a Pleiades catalogue of 111 stars
detected by {\it ROSAT}, with rotation periods $0.26\leq P \leq
9.04$\,d and masses (see below) of $0.33<M/M_{\odot}<0.95$, along with 18
stars with $0.35\leq P \leq 9.41$\,d and $0.18<M/M_{\odot}<0.95$ that have
only X-ray upper limits. The Hartman et al. (2010) catalogue has a
relatively bright magnitude limit and hence few very low-mass stars. 
Nevertheless there were 38 detections and 10 upper limits for objects
classed here as M-dwarfs ($M<0.55\,M_{\odot}$).

Masses were calculated for the field and cluster dwarfs using
bolometric luminosities listed in the source papers and the Z=0.02
Siess et al. (2000) models with (i) a 1\,Gyr age for the field
stars and stars in the Hyades, (ii) a 100\,Myr age for stars in the
Alpha Per and Pleiades clusters, (iii) a 50\,Myr age for stars in the
IC~2391 and IC~2602 clusters.  Rossby numbers were calculated using
equation~3. $L_{x}/L_{\rm bol}$ values were taken from Pizzolato et
al. (2003) and Kiraga \& St{\c e}pie\'n (2007), or from Stauffer et al.
(1994) and Micela et al. (1999) for the Pleiades.  The X-ray fluxes in
these papers are quoted in the 0.1--2.4\,keV {\it ROSAT} passband. The
spectral model adopted for NGC~2547 in section 3.3 predicts that a flux
in the 0.1--2.4\,keV band would only be 6 per cent higher than the
0.3--3\,keV fluxes reported in Table~1. This difference is small enough
to be neglected in our comparisons.

The combined data are shown in Fig.~\ref{comblxlbol},
where correlations between activity and period
and activity and Rossby number are investigated. The stars have been
split into three mass ranges (the same subsets as in section 3.3) 
to see whether period or Rossby number is the best parameter with
which to determine X-ray activity levels across a broad range of
masses and convection zone depths. The numbers of stars taken from each
of the clusters and the field in each of the three mass ranges are:
NGC~2547 (45 for $<0.35\,M_{\odot}$, 25 for $0.35<M/M_{\odot}<0.55$, 26 
for $0.55<M/M_{\odot}<0.95$); IC~2391/2602 (0, 3, 24); Alpha Per (0,
2, 29); Pleiades (10, 38, 71); Hyades (0, 1, 8); Field (8, 27, 44)

The additional data, particularly more fast-rotating K-dwarfs and
slow-rotating M-dwarfs, clarifies a number of issues.
\begin{enumerate}
\item The left hand panels of Fig.~\ref{comblxlbol} show that both
  slowly rotating K- and M-dwarfs have a decreasing level of coronal
  activity at long periods.  However, rotation period alone appears to
  be a poor indicator of X-ray activity for slowly rotating stars.
  K-dwarfs and M-dwarfs follow different rotation-activity
  relationships at long periods -- compare the M-dwarfs with the 
  (red) solid locus which approximately
  indicates the relationship followed by the K-dwarfs. The rotation period
  at which saturation sets in may be longer for lower mass stars.
\item The addition of further fast-rotating K-dwarfs, mainly from other
young clusters, clearly demonstrates a super-saturation effect at
$P<0.3$\,d.  The evidence for super-saturation in M-dwarfs is not convincing:
there are two upper limits (from NGC~2547 and also present in
Fig.~\ref{lxlbolper}) that may suggest super-saturation at $P\simeq
0.2$\,d, in the lowest mass stars. However, the few additional short-period
M-dwarfs from the comparison samples show no indication of
super-saturation down to periods of 0.25\,d.
\item   The Rossby number (or at least the Rossby number found from the convective
  turnover times that we have calculated) unifies the low-activity side
  of the rotation-activity correlation 
  (in the right hand panels of Fig.~\ref{comblxlbol}). The Rossby number is
  capable of predicting the level of X-ray activity with a modest
  scatter and applies to a wide range of masses and
  convection zone depths, including some stars which are fully or
  nearly-fully convective. Coronal saturation occurs, within the limitations of the small
  number statistics for the lowest mass stars, at a similar range of Rossby
  numbers $(-1.8 < \log N_R < -0.8$) for all low-mass stars and at
  a similar value of $L_{x}/L_{\rm bol}$.
\item For $\log N_R < -1.8$ the Rossby number does less well in
  predicting what happens to the coronal activity. Whilst there is
  evidence that some G- and K-dwarfs super-saturate at $\log N_R <
  -1.8$, there is a large scatter. There is no evidence for declining
  activity at low Rossby numbers among the M-dwarfs, unless the two
  upper limits at $\log N_R < -2.5$ in NGC~2547 mark the beginning of
  super-saturation at much lower Rossby numbers. Comparing the left and
  right hand panels of Fig.~\ref{comblxlbol} it may be that, for
  fast-rotating stars, rotation period is the better parameter to
  determine when and if super-saturation occurs.
\item None of these conclusions are affected by the inclusion or
  otherwise of cluster M-dwarfs from the Pizzolato et al. (2003), that
  may be subject to an upward selection bias.
\item There are 3 K-dwarfs (1 detection -- HII\,1095, and 2 upper limits
  -- HII\,370, HII\,793) in the
  Pleiades sample, with $P$ in the range 0.87--0.94\,d, which 
  have anomalously low X-ray activity. We do not believe these are
  examples of coronal super-saturation, but more likely they are
  stars where the period found by Hartman (2010) is a 1-day alias of
  a true longer period. 
\end{enumerate}

\section{Discussion}

A summary of the findings of sections 3 and~4 would be that X-ray
activity rises with decreasing Rossby number, reaches a saturation
plateau at a critical Rossby number ($\log N_R \simeq -0.8$) that is
approximately independent of stellar mass, and then becomes
super-saturated at a smaller Rossby number in a way that {\it is} dependent on
stellar mass -- either in the sense that super-saturation does not occur
at all in low-mass M-dwarfs or if it does, it occurs at much lower Rossby
numbers than for K-dwarfs. Rossby number is the parameter of choice to describe the
occurrence of saturation and the decline of activity at slower rotation
rates, but rotation period may be a better parameter for
predicting the occurrence of super-saturation at $P \sim 0.3$\,d in
K-dwarfs and perhaps $P \sim 0.2$\,d in lower mass stars.

There are a number of ideas that could explain the progression from the
unsaturated to satuarated to super-saturated coronal regimes: (i)
Feedback effects in the dynamo itself may manifest themselves at high
rotation rates. The increasing magnetic field could suppress
differential rotation and the capacity to regenerate poloidal magnetic
field (e.g. Robinson \& Durney 1982; Kitchatinov \& R\"udiger 1993;
Rempel 2006).  (ii) The photosphere may become entirely filled with
equipartion fields (e.g. Reiners, Basri \& Browning 2009). This leads
to saturation of magnetic activity indicators in the corona and
chromosphere. At faster rotation rates it is possible that the magnetic
field emerges in a more restricted way (e.g. constrained to higher
latitudes -- Solanki et al. 1997; St{\c e}pie\'n et al. 2001), reducing
the filling factor again and leading to super-saturation. (iii) At high
rotation rates, centrifugal forces could cause a rise in pressure in
the summits of magnetic loops, which then break open or become
radiatively unstable, reducing the emitting coronal volume and emission
measure (Jardine \& Unruh 1999).

The results from NGC~2547 and from Fig.~\ref{comblxlbol} favour an
explanation for saturation in terms of saturation of the dynamo or
magnetic filling factor, rather than a simple dependence on rotation
rate. The coronal activity of low- and very-low mass 
M-dwarfs in the right hand panels  of Fig.~\ref{comblxlbol} is well
described by the same relationship between activity and Rossby number
as the higher mass K-dwarfs, at least for $\log N_{R}> -1.8$. The Rossby number at which coronal
activity saturates is also broadly similar across a wide range of
masses. These facts suggest that X-ray activity levels depend primarily on Rossby
number, which is a key parameter describing the efficiency of a
magnetic dynamo (Durney \& Robinson 1982; Robinson \& Durney
1982). Supporting evidence for an explanation involving saturation of
magnetic flux generation comes from direct measurements of magnetic
flux in fast rotating M-dwarfs (Saar 1991; Reiners et al. 2009) and
from chromospheric magnetic activity indicators, which also show show
saturation at Rossby numbers of $\simeq 0.1$ (Cardini \&
Cassatella 2007; Marsden, Carter \& Donati 2009). A caveat to this
conclusion is that the convective turnover times and hence Rossby
numbers of the lowest mass stars in our sample are uncertain. In fact
the semi-empirical scaling of $\tau_c \propto L_{\rm bol}^{-1/2}$ was
{\it designed} to minimise the scatter at large Rossby numbers
(Pizzolato et al. 2003). Clearly, better theoretical calculations of $\tau_c$
are desirable for $M<0.5\,M_{\odot}$.

The phenomenon of super-saturation does not seem to be well described
by Rossby numbers calculated using similar $\tau_c$ estimates. Some G- and
K-dwarfs show super-saturation at $\log N_R \simeq -1.8$, but using the
same $\tau_c$ values that tidy up the low-activity side of
Fig.~\ref{comblxlbol} implies that M-dwarfs do not super-saturate
unless at $\log N_R \simeq -2.5$.  This suggests that super-saturation
may not be intrinsic to the dynamo mechanism; a point of view supported
by the lack of super-saturation in the chromospheric emission from very
rapidly rotating G- and K-dwarfs (Marsden, Carter \& Donati 2009). It
is worth noting that the lack of super-saturation in M-dwarfs is
probably not related to any fundamental change in dynamo action, such
as a switch from an interface dynamo to a distributed dynamo as the
convection zone deepens. There are sufficient M-dwarfs in
Fig.~\ref{comblxlbol} with $M>0.35\,M_{\odot}$, which should still have
radiative cores (Siess et al. 2000), to demonstrate that they also show
no signs of super-saturation at the Rossby numbers of super-saturated
G- and K-dwarfs.

\begin{figure*}
\centering
\centering
\begin{minipage}[t]{0.95\textwidth}
\includegraphics[width=80mm]{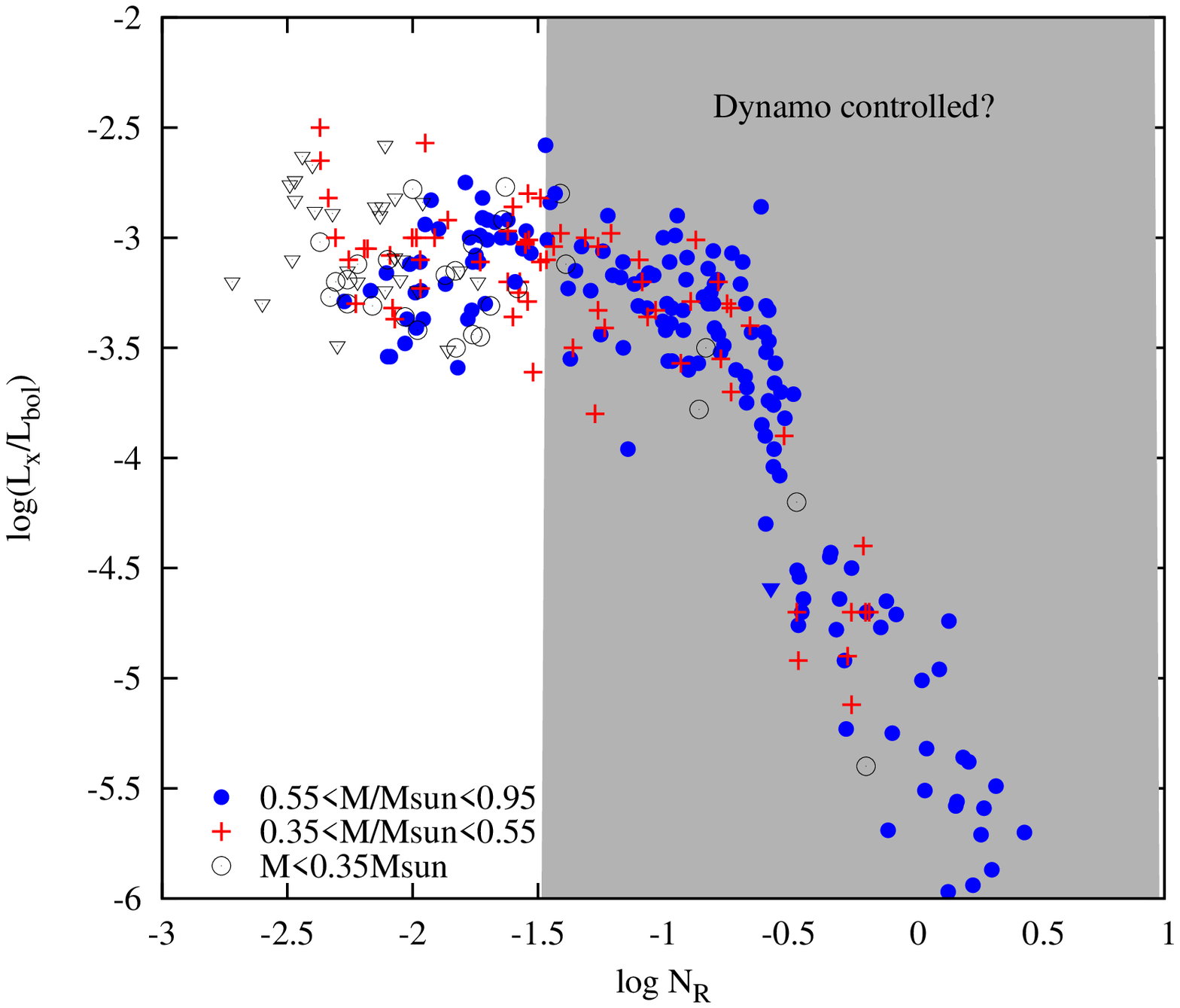}
\includegraphics[width=80mm]{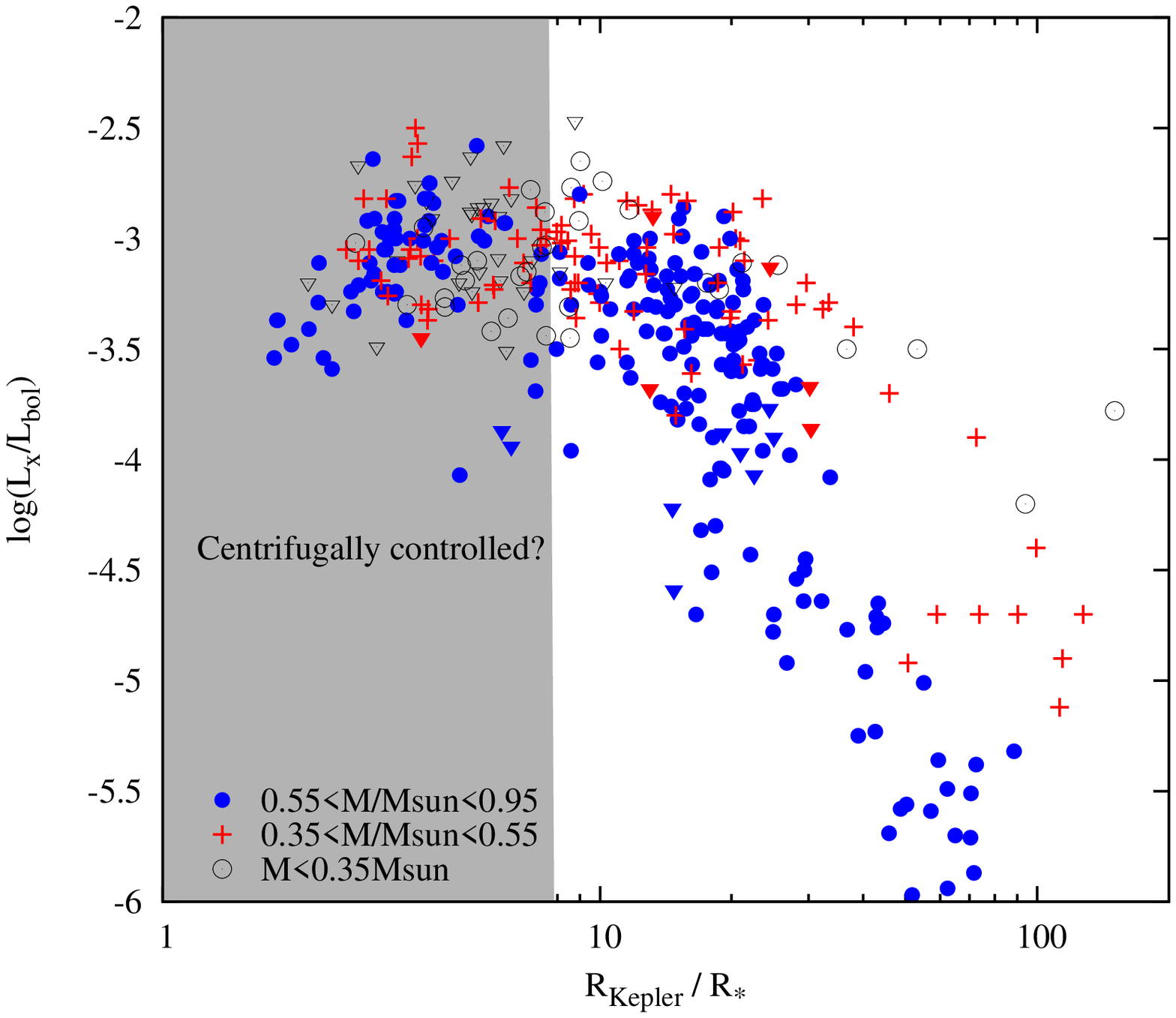}
\end{minipage}
\caption{X-ray activity as a function of Rossby number (left) and Keplerian co-rotation radius
  (right, expressed as a fraction of the stellar radius). The sample is
  divided by mass as in Fig.~\ref{lxlbolper}. The shaded regions
  indicate the approximate regimes in which we hypothesise that the coronal
  activity is controlled by magnetic dynamo efficiency (at large Rossby
  number in the left hand plot) or by centrifugal effects (for small
  Keplerian co-rotation radii in the right hand plot). 
}
\label{rkplot}
\end{figure*}

St{\c e}pie\'n et al. (2001) put forward a hypothesis that a
latitudinal dependence of the heating flux at the base of the
convection zone is caused by the polar dependence of the local gravity
in rapid rotators. This could result in strong poleward updrafts in the
convection zone that sweep magnetic flux tubes to higher latitudes, leaving an
equatorial band that is free from magnetically active regions, hence
reducing the filling factor of magnetically active regions in the
photosphere, chromosphere and corona. In this
model super-saturation occurs when the ratio of centrifugal
acceleration at the surface of the radiative core to the local
gravitational acceleration reaches some critical value $\gamma$. i.e.
\begin{equation}
G\, M_c\, R_c^{-2} = \gamma\, 4\pi^2\, R_c\, P^{-2}\, ,
\end{equation}
where $M_c$ and $R_c$ are the mass and radius of the radiative core.

Leaving aside the issue of what happens in fully convective stars, we
can make the approximation that $M_c R_c^{-3}$ is approximately
proportional to the central density, so that the period $P_{ss}$ at which
super-saturation would be evident depends on central density as 
$P_{\rm ss} \propto \rho_c^{-1/2}$
(assuming that the convection zone rotates as a solid body). The
central density as a function of mass is very time dependent on the
PMS. At 35\,Myr a star of $0.3\,M_{\odot}$ has $\rho_c = 3000$\,
kg\,m$^{-3}$, while a $0.9\,M_{\odot}$ star has $\rho_c =
6900$\,kg\,m$^{-3}$ (Siess et al. 2000).  At 100\,Myr however, the core
of the $0.3\,M_{\odot}$ star is nearly three times denser, while the
$0.9\,M_{\odot}$ star is almost unchanged.  Thanks to the inverse
square root dependence on density however, one would expect super-saturation
to occur at quite similar periods in objects with a range of masses and 
certainly with a variation that is much smaller than if super-saturation occurred at a
fixed Rossby number.  However, there is no evidence of super-saturation
in the chromospheric activity of G- and K-dwarfs which {\em are} coronally
super-saturated (Marsden et al. 2009), and this argues that a simple
restriction of the filling factor due to a polar concentration of the
magnetic field is not the solution.

Jardine \& Unruh (1999) have shown that dynamo saturation or complete
filling by active regions may not be necessary to explain the observed
plateau in X-ray activity and its subsequent decline at very fast
rotation rates. In their model, centrifugal forces act to strip the outer
coronal volume, either because the plasma pressure exceeds what can be
contained by closed magnetic loops (see also Ryan et al. 2005) or
because the coronal plasma becomes radiatively unstable beyond the
Keplerian co-rotation radius (Collier Cameron 1988). The reduced
coronal volume is initially balanced by a rising coronal density, causing
a saturation plateau, but at extreme rotation rates, as more of the
corona is forced open, the X-ray emission measure falls (see also
Jardine 2004). 

We might expect centrifugal effects to become significant in
the most rapidly rotating stars and whilst there will clearly be a
correlation with Rossby number, there will be an important difference in
mass dependence.
The key dimensionless parameter in the centrifugal stripping model is
$\alpha_c$, the co-rotation radius expressed as a multiple of the stellar radius
\begin{equation}
\alpha_c = (G M_{\ast} P^2/ 4 \pi^2 R_{\ast}^{3})^{1/3}  \propto
M_{\ast}^{1/3} R_{\ast}^{-1} P^{2/3}\, ,
\end{equation}
where $P$ is the rotation period. Thus coronal activity should
saturate at some small value of $\alpha_c$ and then super-saturate at
an even smaller $\alpha_c$. In the samples considered here there is a
two order of magnitude range in $P$ but a much smaller
range in $M_{\ast}^{1/3} R_{\ast}^{-1}$.
Hence we expect to see saturation and super-saturation occur at short
periods corresponding to some critical values of $\alpha_c$. However,
in stars with lower masses and smaller radii, these critical $\alpha_c$
values will be reached at {\it shorter} periods, such that $P_{ss}
\propto M^{-1/2} R^{3/2}$, which in NGC~2547 varies from 0.9 (in solar
units) for the most massive stars in our sample to 0.5 in the lowest
mass stars. Hence super-saturation in the M-dwarfs would
occur at shorter periods than for K-dwarfs by a factor approaching 2.

To test these ideas Fig.~\ref{rkplot} shows $L_{x}/L_{\rm bol}$ as a
function of both $\log N_R$ and $\alpha_c$, with the stars grouped into
mass subsets in a similar way to sections~3 and~4. The stellar radii
and masses for the comparison samples were estimated from their
luminosities and the Siess et al. (2000) models as described in
sections~3 and~4.

The $\alpha_c$ parameter, like the rotation period, is a poor predictor
of what happens to X-ray activity in the slowly-rotating and
low-activity regimes. Saturated levels of coronal activity are reached
for $\alpha_c = 10$--30, dependent on the mass of the star. We
interpret this to mean that centrifugal forces have a negligible effect
on coronal structures in this regime. For large values of $\alpha_c$
and concomitantly large values of $\log N_R$ we suppose that coronal
activity is determined by the efficiency of the magnetic dynamo, hence
explaining the reasonably small scatter within the shaded area of the
left hand panel of Fig.~\ref{rkplot}. On the other hand,
super-saturation seems to be achieved when $\alpha_c \la 2.5$ and the
modest scatter within the shaded area of the right hand panel of
Fig.~\ref{rkplot}, compared with that for $\log N_R < -1.8$ in the left
hand panel, suggests that centrifugal effects may start to control the level of
X-ray emission somewhere between this value and $\alpha_c \sim 10$.

In this model it now becomes clear that the reason we have no clear
evidence for super-saturation in M-dwarfs is that they are not
spinning fast enough for their coronae to be affected by centrifugal
forces. There are only two very low-mass M-dwarfs in our sample (from
NGC~2547) with $\alpha_c < 2.5$ and they do have upper limits to their
activity that could be indicative of super-saturation.

If the corona is limited at some small multiple of the co-rotation
radius, this implies that X-ray emitting coronal structures exist up to
this extent above the stellar surface. The arguments for and against
the presence of such extended coronal structures are summarised by
G\"udel (2004). Perhaps the best supporting evidence comes from the
``slingshot prominences'' observed to occur around many rapidly
rotating stars, which often form at or outside the co-rotation
radius (e.g. Collier Cameron \& Robinson 1989; Jeffries 1993; Barnes et
al. 1998), which may signal the action of the centrifugal stripping process.

\section{Summary}

We have determined the level of X-ray activity (in terms of
$L_{x}/L_{\rm bol}$) for a sample of low-mass K- and M-dwarfs with
rotation periods of 0.2--10\,d, in the young ($\simeq 35$\,Myr) open
cluster NGC~2547. A deep {\it XMM-Newton} observation is able to
detect X-rays from almost all stars with $M\geq 0.35\,M_{\odot}$ and
provides detections or upper limits for many more with lower masses.
Most targets exhibit saturated levels of X-ray activity ($L_{x}/L_{\rm
bol}\simeq 10^{-3}$), but a few of the most rapidly rotating showed
evidence of lower, ``super-saturated'' activity. The evidence for
super-saturation in M-dwarfs with $M<0.55\,M_{\odot}$ is weak, limited
to three objects with short periods and very small Rossby numbers,
which have 3-sigma upper limits to their coronal activity of $L_{x}/L_{\rm
bol}\leq 10^{-3.3}$ to $10^{-3.4}$.

These data were combined with X-ray measurements of fast-rotating low-mass
stars in the field and from other young clusters and the results were considered
in terms of stellar rotation period and in terms of the Rossby
number, which is thought to be diagnostic of magnetic dynamo
efficiency.  The main result is that while coronal saturation appears
to occur below a threshold Rossby number, $\log N_R \la -0.8$,
independent of stellar mass, there is evidence that super-saturation
does not occur at a fixed Rossby number. Super-saturation in the
lowest mass M-dwarfs occurs, if at all, at Rossby numbers that are at least three
times smaller than those for which super-saturation is observed to
occur in G- and K-dwarfs; there is no evidence for super-saturation in
M-dwarfs with $\log N_R > -2.3$.  Instead it appears that a rotation period
$<0.3$\,d is a more accurate predictor of when super-saturation
commences and there are only a few M-dwarfs rotating at these periods in
our sample.

A caveat to this result is that our calculated Rossby numbers rely on
rather uncertain semi-empirical values of the convective turnover
time. However, these turnover times do unify the slowly-rotating
side of the rotation-activity relation in the sense that the use of
Rossby numbers significantly reduces the scatter in this
relationship and predicts a common threshold Rossby number for coronal
saturation.

These phenomena can be interpreted within
a framework where coronal saturation occurs due to saturation of the
dynamo itself at a critical Rossby number, or perhaps due to complete
filling of the surface by magnetically active regions. Coronal
super-saturation is probably not an intrinsic property of the dynamo,
but instead associated with topological changes in the coronal magnetic
field with fast rotation rates. The observations favour the centrifugal
stripping scenario of Jardine \& Unruh (1999) and Jardine (2004), in
which the reduction of the available coronal volume due to
instabilities induced by centrifugal forces leads to a drop in X-ray
emission as the Keplerian co-rotation radius moves in towards the
surface of the star.

If centrifugal stripping is the correct explanation for
super-saturation then M-dwarfs should super-saturate at shorter periods
than K-dwarfs, by factors of up to $\sim 2$. The data analysed here are
sparse but consistent with this idea. Determining the X-ray emission
from a small sample of rapidly rotating ($P<0.25$\,d) M-dwarfs would
resolve this issue. The centrifugal stripping idea may also explain the
positive correlation between X-ray activity and rotation period seen in
very young PMS stars by Stassun et al. (2004) and Preibisch et
al. (2005). It could be that the fastest rotating young PMS stars are
centrifugally stripped whilst those of slower rotation are merely
saturated. Investigating this in detail faces formidable observational
and theoretical difficulties, not least the estimation of the
intrinsic, non-accretion related X-ray emission, individual absorption
column densities for stars and the determination of PMS stellar radii,
masses and convective turnover times.

\section*{Acknowledgements}
Based on observations obtained with XMM-Newton, an ESA science mission
with instruments and contributions directly funded by ESA Member States
and NASA. RJJ acknowledges receipt of a Science and
Technology Facilities Council postgraduate studentship.

\nocite{barnes98}
\nocite{jeffries93}
\nocite{collier89b}
\nocite{jeffries98n2547}
\nocite{ryan05}
\nocite{collier88}
\nocite{robinson82}
\nocite{rempel06}
\nocite{reinersbsat09}
\nocite{pallavicini81a}
\nocite{mangeney84}
\nocite{noyes84}
\nocite{dobson89}
\nocite{pizzolato03}
\nocite{gudel04}
\nocite{gilliland86}
\nocite{kim96}
\nocite{vilhu87sat}
\nocite{stauffer94}
\nocite{stepien94}
\nocite{randich96alphaper}
\nocite{patten96}
\nocite{kiraga07}
\nocite{doyle96}
\nocite{solanki97}
\nocite{jardine99}
\nocite{stauffer97ic23912602}
\nocite{kitchatinov94}
\nocite{stepien01}
\nocite{jardine04}
\nocite{jeffries06}
\nocite{james00}
\nocite{irwin08}
\nocite{jeffries05}
\nocite{naylor06}
\nocite{jeffries04}
\nocite{struder01}
\nocite{turner01}
\nocite{dantona97}
\nocite{claria82}
\nocite{guainazzi10}
\nocite{lyra06}
\nocite{mekal95}
\nocite{garcia08}
\nocite{briggs03}
\nocite{saxton03}
\nocite{bohlin78}
\nocite{leggett96}
\nocite{carrera07}
\nocite{telleschi05}
\nocite{siess00}
\nocite{mayne08}
\nocite{ventura98}
\nocite{randich95ic2602}
\nocite{voges99}
\nocite{huensch99}
\nocite{randichsupersat98}
\nocite{durney82}
\nocite{saar91}
\nocite{marsden09}
\nocite{cardini07}
\nocite{prosser96}
\nocite{marino03a}
\nocite{gagnepleiades95}
\nocite{cruddace84}
\nocite{audard00}
\nocite{hartman10}
\nocite{naylor02}
\nocite{kraft91}
\nocite{park06}
\nocite{stassun04b}
\nocite{preibisch05b}
\nocite{dahm07}
\nocite{landin10}
\nocite{perryman98}
\nocite{stauffer99}
\nocite{stauffer98}
\nocite{barrado04}

\bibliographystyle{mn2e} 
\bibliography{iau_journals,master}

\begin{thebibliography}{}

\bibitem[\protect\citeauthoryear{{Audard}, {G{\"u}del}, {Drake} \&
  {Kashyap}}{{Audard} et~al.}{2000}]{audard00}
{Audard} M.,  {G{\"u}del} M.,  {Drake} J.~J.,    {Kashyap} V.~L.,  2000, ApJ,
  541, 396

\bibitem[\protect\citeauthoryear{Barnes, Collier-Cameron, Unruh, Donati \&
  Hussain}{Barnes et~al.}{1998}]{barnes98}
Barnes J.~R.,  Collier-Cameron A.,  Unruh Y.~C.,  Donati J.~F.,    Hussain G.
  A.~J.,  1998, MNRAS, 299, 904

\bibitem[\protect\citeauthoryear{{Barrado y Navascu{\'e}s}, {Stauffer} \&
  {Jayawardhana}}{{Barrado y Navascu{\'e}s} et~al.}{2004}]{barrado04}
{Barrado y Navascu{\'e}s} D.,  {Stauffer} J.~R.,    {Jayawardhana} R.,  2004,
  ApJ, 614, 386

\bibitem[\protect\citeauthoryear{Bohlin, Savage \& Drake}{Bohlin
  et~al.}{1978}]{bohlin78}
Bohlin R.~C.,  Savage B.~D.,    Drake J.~F.,  1978, ApJ, 224, 132

\bibitem[\protect\citeauthoryear{Briggs \& Pye}{Briggs \& Pye}{2003}]{briggs03}
Briggs K.,  Pye J.~P.,  2003, MNRAS, 345, 714

\bibitem[\protect\citeauthoryear{{Cardini} \& {Cassatella}}{{Cardini} \&
  {Cassatella}}{2007}]{cardini07}
{Cardini} D.,  {Cassatella} A.,  2007, \apj, 666, 393

\bibitem[\protect\citeauthoryear{{Carrera}, {Ebrero}, {Mateos}, {Ceballos},
  {Corral}, {Barcons}, {Page}, {Rosen}, {Watson}, {Tedds}, {Della Ceca},
  {Maccacaro}, {Brunner}, {Freyberg}, {Lamer}, {Bauer} \& {Ueda}}{{Carrera}
  et~al.}{2007}]{carrera07}
{Carrera} F.~J.,  {Ebrero} J.,  {Mateos} S.,  {Ceballos} M.~T.,  {Corral} A.,
  {Barcons} X.,  {Page} M.~J.,  {Rosen} S.~R.,  {Watson} M.~G.,  {Tedds} J.~A.,
   {Della Ceca} R.,  {Maccacaro} T.,  {Brunner} H.,  {Freyberg} M.,  {Lamer}
  G.,  {Bauer} F.~E.,    {Ueda} Y.,  2007, A\&A, 469, 27

\bibitem[\protect\citeauthoryear{Clari\'{a}}{Clari\'{a}}{1982}]{claria82}
Clari\'{a} J.~J.,  1982, A\&AS, 47, 323

\bibitem[\protect\citeauthoryear{Collier-Cameron}{Collier-Cameron}{1988}]{coll%
ier88}
Collier-Cameron A.,  1988, MNRAS, 233, 235

\bibitem[\protect\citeauthoryear{Collier-Cameron \& Robinson}{Collier-Cameron
  \& Robinson}{1989}]{collier89b}
Collier-Cameron A.~C.,  Robinson R.~D.,  1989, MNRAS, 238, 657

\bibitem[\protect\citeauthoryear{{Cruddace} \& {Dupree}}{{Cruddace} \&
  {Dupree}}{1984}]{cruddace84}
{Cruddace} R.~G.,  {Dupree} A.~K.,  1984, ApJ, 277, 263

\bibitem[\protect\citeauthoryear{{Dahm}, {Simon}, {Proszkow} \&
  {Patten}}{{Dahm} et~al.}{2007}]{dahm07}
{Dahm} S.~E.,  {Simon} T.,  {Proszkow} E.~M.,    {Patten} B.~M.,  2007, AJ,
  134, 999

\bibitem[\protect\citeauthoryear{D'Antona \& Mazzitelli}{D'Antona \&
  Mazzitelli}{1997}]{dantona97}
D'Antona F.,  Mazzitelli I.,  1997, Mem. Soc. Astr. It., 68, 807

\bibitem[\protect\citeauthoryear{Dobson \& Radick}{Dobson \&
  Radick}{1989}]{dobson89}
Dobson A.~K.,  Radick R.~R.,  1989, ApJ, 344, 907

\bibitem[\protect\citeauthoryear{{Doyle}}{{Doyle}}{1996}]{doyle96}
{Doyle} J.~G.,  1996, A\&A, 307, L45

\bibitem[\protect\citeauthoryear{Durney \& Robinson}{Durney \&
  Robinson}{1982}]{durney82}
Durney B.~R.,  Robinson R.~D.,  1982, ApJ, 253, 290

\bibitem[\protect\citeauthoryear{Gagn\'{e}, Caillault \& Stauffer}{Gagn\'{e}
  et~al.}{1995}]{gagnepleiades95}
Gagn\'{e} M.,  Caillault J.~P.,    Stauffer J.~R.,  1995, ApJ, 450, 217

\bibitem[\protect\citeauthoryear{{Garc{\'{\i}}a-Alvarez}, {Drake}, {Kashyap},
  {Lin} \& {Ball}}{{Garc{\'{\i}}a-Alvarez} et~al.}{2008}]{garcia08}
{Garc{\'{\i}}a-Alvarez} D.,  {Drake} J.~J.,  {Kashyap} V.~L.,  {Lin} L.,
  {Ball} B.,  2008, \apj, 679, 1509

\bibitem[\protect\citeauthoryear{Gilliland}{Gilliland}{1986}]{gilliland86}
Gilliland R.~L.,  1986, ApJ, 300, 339

\bibitem[\protect\citeauthoryear{Guainazzi}{Guainazzi}{2010}]{guainazzi10}
Guainazzi M.,  2010, Technical Report XMM-SOC-CAL-TN-0018, EPIC status of
  calibration and data analysis, Version 2.9.
XMM-Newton SOC

\bibitem[\protect\citeauthoryear{G{\" u}del}{G{\" u}del}{2004}]{gudel04}
G{\" u}del M.,  2004, ARA\&A, 12, 71

\bibitem[\protect\citeauthoryear{{Hartman}, {Bakos}, {Kov{\'a}cs} \&
  {Noyes}}{{Hartman} et~al.}{2010}]{hartman10}
{Hartman} J.~D.,  {Bakos} G.~{\'A}.,  {Kov{\'a}cs} G.,    {Noyes} R.~W.,  2010,
  ArXiv e-prints

\bibitem[\protect\citeauthoryear{{H{\"u}nsch}, {Schmitt}, {Sterzik} \&
  {Voges}}{{H{\"u}nsch} et~al.}{1999}]{huensch99}
{H{\"u}nsch} M.,  {Schmitt} J.~H.~M.~M.,  {Sterzik} M.~F.,    {Voges} W.,
  1999, \aaps, 135, 319

\bibitem[\protect\citeauthoryear{{Irwin}, {Hodgkin}, {Aigrain}, {Bouvier},
  {Hebb} \& {Moraux}}{{Irwin} et~al.}{2008}]{irwin08}
{Irwin} J.,  {Hodgkin} S.,  {Aigrain} S.,  {Bouvier} J.,  {Hebb} L.,
  {Moraux} E.,  2008, \mnras, 383, 1588

\bibitem[\protect\citeauthoryear{{James}, {Jardine}, {Jeffries}, {Randich},
  {Collier Cameron} \& {Ferreira}}{{James} et~al.}{2000}]{james00}
{James} D.~J.,  {Jardine} M.~M.,  {Jeffries} R.~D.,  {Randich} S.,  {Collier
  Cameron} A.,    {Ferreira} M.,  2000, \mnras, 318, 1217

\bibitem[\protect\citeauthoryear{{Jardine}}{{Jardine}}{2004}]{jardine04}
{Jardine} M.,  2004, A\&A, 414, L5

\bibitem[\protect\citeauthoryear{Jardine \& Unruh}{Jardine \&
  Unruh}{1999}]{jardine99}
Jardine M.,  Unruh Y.~C.,  1999, A\&A, 346, 883

\bibitem[\protect\citeauthoryear{Jeffries}{Jeffries}{1993}]{jeffries93}
Jeffries R.~D.,  1993, MNRAS, 262, 369

\bibitem[\protect\citeauthoryear{{Jeffries}, {Evans}, {Pye} \&
  {Briggs}}{{Jeffries} et~al.}{2006}]{jeffries06}
{Jeffries} R.~D.,  {Evans} P.~A.,  {Pye} J.~P.,    {Briggs} K.~R.,  2006,
  MNRAS, 367, 781

\bibitem[\protect\citeauthoryear{Jeffries, Naylor, Devey \& Totten}{Jeffries
  et~al.}{2004}]{jeffries04}
Jeffries R.~D.,  Naylor T.,  Devey C.~R.,    Totten E.~J.,  2004, MNRAS, 351,
  1401

\bibitem[\protect\citeauthoryear{Jeffries \& Oliveira}{Jeffries \&
  Oliveira}{2005}]{jeffries05}
Jeffries R.~D.,  Oliveira J.~M.,  2005, MNRAS, 358, 13

\bibitem[\protect\citeauthoryear{Jeffries \& Tolley}{Jeffries \&
  Tolley}{1998}]{jeffries98n2547}
Jeffries R.~D.,  Tolley A.~J.,  1998, MNRAS, 300, 331

\bibitem[\protect\citeauthoryear{Kim \& Demarque}{Kim \&
  Demarque}{1996}]{kim96}
Kim Y.-C.,  Demarque P.,  1996, ApJ, 457, 340

\bibitem[\protect\citeauthoryear{{Kiraga} \& {Stepien}}{{Kiraga} \&
  {Stepien}}{2007}]{kiraga07}
{Kiraga} M.,  {Stepien} K.,  2007, Acta Astronomica, 57, 149

\bibitem[\protect\citeauthoryear{{Kitchatinov}, {Ruediger} \&
  {Kueker}}{{Kitchatinov} et~al.}{1994}]{kitchatinov94}
{Kitchatinov} L.~L.,  {Ruediger} G.,    {Kueker} M.,  1994, \aap, 292, 125

\bibitem[\protect\citeauthoryear{{Kraft}, {Burrows} \& {Nousek}}{{Kraft}
  et~al.}{1991}]{kraft91}
{Kraft} R.~P.,  {Burrows} D.~N.,    {Nousek} J.~A.,  1991, ApJ, 374, 344

\bibitem[\protect\citeauthoryear{{Landin}, {Mendes} \& {Vaz}}{{Landin}
  et~al.}{2010}]{landin10}
{Landin} N.~R.,  {Mendes} L.~T.~S.,    {Vaz} L.~P.~R.,  2010, A\&A, 510, A46+

\bibitem[\protect\citeauthoryear{Leggett, Allard, Berriman, Dahn \&
  Hauschildt}{Leggett et~al.}{1996}]{leggett96}
Leggett S.~K.,  Allard F.,  Berriman G.,  Dahn C.~C.,    Hauschildt P.~H.,
  1996, ApJS, 104, 117

\bibitem[\protect\citeauthoryear{{Lyra}, {Moitinho}, {van der Bliek} \&
  {Alves}}{{Lyra} et~al.}{2006}]{lyra06}
{Lyra} W.,  {Moitinho} A.,  {van der Bliek} N.~S.,    {Alves} J.,  2006, \aap,
  453, 101

\bibitem[\protect\citeauthoryear{{Mangeney} \& {Praderie}}{{Mangeney} \&
  {Praderie}}{1984}]{mangeney84}
{Mangeney} A.,  {Praderie} F.,  1984, A\&A, 130, 143

\bibitem[\protect\citeauthoryear{Marino, Micela, Peres \& Sciortino}{Marino
  et~al.}{2003}]{marino03a}
Marino A.,  Micela G.,  Peres G.,    Sciortino S.,  2003, A\&A, 406, 629

\bibitem[\protect\citeauthoryear{{Marsden}, {Carter} \& {Donati}}{{Marsden}
  et~al.}{2009}]{marsden09}
{Marsden} S.~C.,  {Carter} B.~D.,    {Donati} J.,  2009, \mnras, 399, 888

\bibitem[\protect\citeauthoryear{{Mayne} \& {Naylor}}{{Mayne} \&
  {Naylor}}{2008}]{mayne08}
{Mayne} N.~J.,  {Naylor} T.,  2008, \mnras, 386, 261

\bibitem[\protect\citeauthoryear{Mewe, Kaastra \& Leidahl}{Mewe
  et~al.}{1995}]{mekal95}
Mewe R.,  Kaastra J.~S.,    Leidahl D.~A.,  1995, Technical report, The HEASARC
  Journal, Legacy Volume 6.
Goddard Space Flight Center

\bibitem[\protect\citeauthoryear{Naylor \& Jeffries}{Naylor \&
  Jeffries}{2006}]{naylor06}
Naylor T.,  Jeffries R.~D.,  2006, MNRAS, 373, 1251

\bibitem[\protect\citeauthoryear{Naylor, Totten, Jeffries, Pozzo, Devey \&
  Thompson}{Naylor et~al.}{2002}]{naylor02}
Naylor T.,  Totten E.~J.,  Jeffries R.~D.,  Pozzo M.,  Devey C.~R.,    Thompson
  S.~A.,  2002, MNRAS, 335, 291

\bibitem[\protect\citeauthoryear{Noyes, Hartmann, Baliunas, Duncan \&
  Vaughan}{Noyes et~al.}{1984}]{noyes84}
Noyes R.~W.,  Hartmann L.,  Baliunas S.~L.,  Duncan D.~K.,    Vaughan A.~H.,
  1984, ApJ, 279, 763

\bibitem[\protect\citeauthoryear{Pallavicini, Golub, Rosner, Vaiana, Ayres \&
  Linsky}{Pallavicini et~al.}{1981}]{pallavicini81a}
Pallavicini R.,  Golub L.,  Rosner R.,  Vaiana G.~S.,  Ayres T.,    Linsky
  J.~L.,  1981, ApJ, 248, 279

\bibitem[\protect\citeauthoryear{{Park}, {Kashyap}, {Siemiginowska}, {van Dyk},
  {Zezas}, {Heinke} \& {Wargelin}}{{Park} et~al.}{2006}]{park06}
{Park} T.,  {Kashyap} V.~L.,  {Siemiginowska} A.,  {van Dyk} D.~A.,  {Zezas}
  A.,  {Heinke} C.,    {Wargelin} B.~J.,  2006, ApJ, 652, 610

\bibitem[\protect\citeauthoryear{Patten \& Simon}{Patten \&
  Simon}{1996}]{patten96}
Patten B.~M.,  Simon T.,  1996, ApJS, 106, 489

\bibitem[\protect\citeauthoryear{{Perryman}, {Brown}, {Lebreton}, {Gomez},
  {Turon}, {Cayrel de Strobel}, {Mermilliod}, {Robichon}, {Kovalevsky} \&
  {Crifo}}{{Perryman} et~al.}{1998}]{perryman98}
{Perryman} M.~A.~C.,  {Brown} A.~G.~A.,  {Lebreton} Y.,  {Gomez} A.,  {Turon}
  C.,  {Cayrel de Strobel} G.,  {Mermilliod} J.~C.,  {Robichon} N.,
  {Kovalevsky} J.,    {Crifo} F.,  1998, A\&A, 331, 81

\bibitem[\protect\citeauthoryear{Pizzolato, Maggio, Micela, Sciortino \&
  Ventura}{Pizzolato et~al.}{2003}]{pizzolato03}
Pizzolato N.,  Maggio A.,  Micela G.,  Sciortino S.,    Ventura P.,  2003,
  A\&A, 397, 147

\bibitem[\protect\citeauthoryear{Preibisch, Kim, Favata, Feigelson, Flaccomio,
  Getman, Micela, Sciortino, Stassun, Stelzer \& Zinnecker}{Preibisch
  et~al.}{2005}]{preibisch05b}
Preibisch T.,  Kim Y.~C.,  Favata F.,  Feigelson E.~D.,  Flaccomio E.,  Getman
  K.,  Micela G.,  Sciortino S.,  Stassun K.,  Stelzer B.,    Zinnecker H.,
  2005, ApJS, 160, 401

\bibitem[\protect\citeauthoryear{Prosser, Randich, Stauffer, Schmitt \&
  Simon}{Prosser et~al.}{1996}]{prosser96}
Prosser C.~F.,  Randich S.,  Stauffer J.~R.,  Schmitt J. H. M.~M.,    Simon T.,
   1996, AJ, 112, 1570

\bibitem[\protect\citeauthoryear{{Randich}}{{Randich}}{1998}]{randichsupersat9%
8}
{Randich} S.,  1998, in {R.~A.~Donahue \& J.~A.~Bookbinder} ed., Cool Stars,
  Stellar Systems, and the Sun Vol.~154 of Astronomical Society of the Pacific
  Conference Series, {Supersaturation in X-ray Emission of Cluster Stars}.
pp 501--+

\bibitem[\protect\citeauthoryear{Randich, Schmitt, Prosser \& Stauffer}{Randich
  et~al.}{1995}]{randich95ic2602}
Randich S.,  Schmitt J. H. M.~M.,  Prosser C.~F.,    Stauffer J.~R.,  1995,
  A\&A, 300, 134

\bibitem[\protect\citeauthoryear{Randich, Schmitt, Prosser \& Stauffer}{Randich
  et~al.}{1996}]{randich96alphaper}
Randich S.,  Schmitt J. H. M.~M.,  Prosser C.~F.,    Stauffer J.~R.,  1996,
  A\&A, 305, 785

\bibitem[\protect\citeauthoryear{{Reiners}, {Basri} \& {Browning}}{{Reiners}
  et~al.}{2009}]{reinersbsat09}
{Reiners} A.,  {Basri} G.,    {Browning} M.,  2009, \apj, 692, 538

\bibitem[\protect\citeauthoryear{{Rempel}}{{Rempel}}{2006}]{rempel06}
{Rempel} M.,  2006, \apj, 647, 662

\bibitem[\protect\citeauthoryear{{Robinson} \& {Durney}}{{Robinson} \&
  {Durney}}{1982}]{robinson82}
{Robinson} R.~D.,  {Durney} B.~R.,  1982, A\&A, 108, 322

\bibitem[\protect\citeauthoryear{{Ryan}, {Neukirch} \& {Jardine}}{{Ryan}
  et~al.}{2005}]{ryan05}
{Ryan} R.~D.,  {Neukirch} T.,    {Jardine} M.,  2005, A\&A, 433, 323

\bibitem[\protect\citeauthoryear{Saar}{Saar}{1991}]{saar91}
Saar S.~H.,  1991, in Tuominen I.,  Moss D.,   {R\"udiger} G.,  eds, {The Sun
  and Cool Stars: Activity, Magnetism, Dynamos} {Recent advances in the
  observation and analysis of stellar magnetic fields}.
Springer, Berlin, p.~389

\bibitem[\protect\citeauthoryear{Saxton}{Saxton}{2003}]{saxton03}
Saxton R.~D.,  2003, Technical Report XMM-SOC-CAL-TN-0023, A statistical
  evaluation of the EPIC flux calibration, Version 2.0.
XMM-Newton SOC

\bibitem[\protect\citeauthoryear{Siess, Dufour \& Forestini}{Siess
  et~al.}{2000}]{siess00}
Siess L.,  Dufour E.,    Forestini M.,  2000, A\&A, 358, 593

\bibitem[\protect\citeauthoryear{{Solanki}, {Motamen} \& {Keppens}}{{Solanki}
  et~al.}{1997}]{solanki97}
{Solanki} S.~K.,  {Motamen} S.,    {Keppens} R.,  1997, A\&A, 324, 943

\bibitem[\protect\citeauthoryear{{Stassun}, {Ardila}, {Barsony}, {Basri} \&
  {Mathieu}}{{Stassun} et~al.}{2004}]{stassun04b}
{Stassun} K.~G.,  {Ardila} D.~R.,  {Barsony} M.,  {Basri} G.,    {Mathieu}
  R.~D.,  2004, AJ, 127, 3537

\bibitem[\protect\citeauthoryear{Stauffer, Barrado~y Navascu\'{e}s, Bouvier,
  Morrison, Harding, Luhman, Stanke, McCaughrean, Terndrup, Allen \&
  Assouad}{Stauffer et~al.}{1999}]{stauffer99}
Stauffer J.~R.,  Barrado~y Navascu\'{e}s D.,  Bouvier J.,  Morrison H.~L.,
  Harding P.,  Luhman K.,  Stanke T.,  McCaughrean M.,  Terndrup D.~M.,  Allen
  L.,    Assouad P.,  1999, ApJ, 527, 219

\bibitem[\protect\citeauthoryear{Stauffer, Caillault, Gagn\'{e}, Prosser \&
  Hartmann}{Stauffer et~al.}{1994}]{stauffer94}
Stauffer J.~R.,  Caillault J.~P.,  Gagn\'{e} M.,  Prosser C.~F.,    Hartmann
  L.~W.,  1994, ApJS, 91, 625

\bibitem[\protect\citeauthoryear{Stauffer, Hartmann, Prosser, Randich,
  Balachandran, Patten, Simon \& Giampapa}{Stauffer
  et~al.}{1997}]{stauffer97ic23912602}
Stauffer J.~R.,  Hartmann L.~W.,  Prosser C.~F.,  Randich S.,  Balachandran S.,
   Patten B.~M.,  Simon T.,    Giampapa M.,  1997, ApJ, 479, 776

\bibitem[\protect\citeauthoryear{Stauffer, Schultz \& Kirkpatrick}{Stauffer
  et~al.}{1998}]{stauffer98}
Stauffer J.~R.,  Schultz G.,    Kirkpatrick J.~D.,  1998, ApJ, 499, L199

\bibitem[\protect\citeauthoryear{{St{\c e}pie{\'n}}, {Schmitt} \&
  {Voges}}{{St{\c e}pie{\'n}} et~al.}{2001}]{stepien01}
{St{\c e}pie{\'n}} K.,  {Schmitt} J.~H.~M.~M.,    {Voges} W.,  2001, \aap, 370,
  157

\bibitem[\protect\citeauthoryear{{Stepien}}{{Stepien}}{1994}]{stepien94}
{Stepien} K.,  1994, A\&A, 292, 191

\bibitem[\protect\citeauthoryear{{Str\"uder, L. et al.}}{{Str\"uder, L. et
  al.}}{2001}]{struder01}
{Str\"uder, L. et al.} 2001, A\&A, 365, L18

\bibitem[\protect\citeauthoryear{Telleschi, Gudel, Briggs, Audard, Ness \&
  Skinner}{Telleschi et~al.}{2005}]{telleschi05}
Telleschi A.,  Gudel M.,  Briggs K.,  Audard M.,  Ness J.~U.,    Skinner S.~L.,
   2005, ApJ, 622, 653

\bibitem[\protect\citeauthoryear{{Turner, M. J. L. et al.}}{{Turner, M. J. L.
  et al.}}{2001}]{turner01}
{Turner, M. J. L. et al.} 2001, A\&A, 365, L27

\bibitem[\protect\citeauthoryear{Ventura, Zeppieri, Mazzitelli \&
  D'Antona}{Ventura et~al.}{1998}]{ventura98}
Ventura P.,  Zeppieri A.,  Mazzitelli I.,    D'Antona F.,  1998, A\&A, 331,
  1011

\bibitem[\protect\citeauthoryear{Vilhu \& Walter}{Vilhu \&
  Walter}{1987}]{vilhu87sat}
Vilhu O.,  Walter F.~M.,  1987, ApJ, 321, 958

\bibitem[\protect\citeauthoryear{Voges}{Voges}{1999}]{voges99}
Voges W. e.~a.,  1999, A\&A, 349, 389

\end{thebibliography}


\bsp 

\label{lastpage}

\end{document}